\documentclass[useAMS,usenatbib]{mn2e}
\usepackage{graphicx,natbib}
\usepackage{psfrag}
\usepackage{amssymb}
\usepackage{float}

\def\aV{\mbox{$\rm A_V$}}

\def\jh{\mbox{$(J-H)$}}
\def\hk{\mbox{$(H-K_s)$}}

\def\ms{\mbox{$M_\odot$}}

\title[Galactic anticentre star clusters]{Towards a census of the Galactic anticentre star clusters II: exploring lower overdensities}

\author[D. Camargo, C. Bonatto and E. Bica]{D. Camargo$^1$, C. Bonatto$^1$ and E. Bica$^1$\\
$^1$ Departamento de Astronomia, Universidade Federal do Rio Grande do Sul, 
Av. Bento Gon\c{c}alves 9500\\
Porto Alegre 91501-970, RS, Brazil}

\begin{document}

\pagerange{\pageref{firstpage}--\pageref{lastpage}}

\maketitle

\label{firstpage}

\begin{abstract}
We investigate the nature of 48 low-level stellar overdensities (from Froebrich, Scholz,
and Raftery catalogue - FSR07) projected towards the Galactic anticentre and derive
fundamental parameters for the confirmed clusters, thus improving the open cluster
(OC) census in that direction. Parameters are derived with field-star decontaminated
photometry, colour-magnitude filters and stellar radial density profiles. Among the 48
targets, we identified 18 star clusters, 6 previously studied OCs, and 7 probable
clusters that require deeper photometry to establish the nature. We discovered 7 new
clusters, 6 of them forming an association of clusters with BPI 14, FSR 777, Kronberger 1, and Stock 8 in the region of the nebula IC 417 and related to the Aur OB2
association, and one embedded in the nebula Sh2-229. We also derive parameters for
these three non-FSR07 clusters, because they are important in determining the structure of the Galactic anticentre. Thus, 58 objects are analysed in this work and we
could derive fundamental parameters for 28 of them. The scenario in the IC
417 star forming region is consistent with a sequential event. FSR 888 and FSR 890
are embedded in Sh2-249 within the Gem OB1 association. According to the distance
derived for these clusters and those in the association of clusters, both Aur OB2 and
Gem OB1 are located in the Perseus arm.
\end{abstract}


\section{Introduction}
\label{Intro}

Star clusters are often considered as building blocks of galaxies. Understanding how these objects form and evolve is vital to our comprehension of the structure, formation and evolution of galaxies. For instance, the open cluster (OC) system has been used to analyse the structure, dynamics, composition, and evolution of the Galactic disk \citep{Friel95, Bonatto06a, Piskunov06}. Young OCs, in particular, have been used as tracers of the spiral pattern of the Galaxy \citep[][and references therein]{Moffat79, Bobylev07, Vazquez08}. In practical terms, astrophysical parameters can in fact be determined for an OC more easily than for a single star.

\begin{table}
\centering
{\footnotesize
\caption{General data on the FSR star cluster candidates.}
\label{tab1}
\renewcommand{\tabcolsep}{0.98mm}
\renewcommand{\arraystretch}{1.1}
\begin{tabular}{lrrrrrrrr}
\hline
\hline
Target&$\alpha(2000)$&$\delta(2000)$&$\ell$&$b$&$R_C$&$R_t$&$Q$\\
&(h\,m\,s)&$(^{\circ}\,^{\prime}\,^{\prime\prime})$&$(^{\circ})$&$(^{\circ})$&$(')$&$(')$& \\
($1$)&($2$)&($3$)&($4$)&($5$)&($6$)&($7$)&($8$)\\

\hline
FSR 707 &5:16:06&47:37:24&161.197&5.421&0.009&0.441&5\\
FSR 716 &5:11:10&45:42:46&162.259&3.619&0.043&0.128&4\\
FSR 722 &5:01:49&44:07:42&162.532&1.346&0.006&0.169&4\\
FSR 734  &5:03:22&42:24:49&164.062&0.522&0.011&0.074&4\\
FSR 738 &3:57:34&28:37:02&165.169&-18.602&0.012&0.594&4\\
FSR 746 &6:20:44&46:48:27&167.421&14.543&0.007&0.026&4\\
FSR 759  &5:01:10&35:47:41&169.042&-3.863&0.007&0.053&4\\
FSR 761 &5:33:23&39:50:44&169.414&3.665&0.007&0.375&4\\
FSR 763 &5:34:39&39:08:12&170.145&3.489&0.021&0.085&4\\
FSR 768 &4:39:52&29:44:22&170.933&-11.153&0.071&0.142&5\\
FSR 771 &5:03:47&32:08:30&172.285&-5.651&0.061&0.307&5\\
FSR 777 &5:27:31&34:44:01&173.047&-0.118&0.056&0.112&4\\
FSR 780 &5:27:26&34:24:12&173.313&-0.314&0.007&0.078&5\\
FSR 784 &5:40:48&35:55:06&173.521&2.8&0.014&0.071&4\\
FSR 798 &4:31:39&22:39:36&175.267&-17.142&0.014&0.708&4\\
FSR 799 &5:42:20&33:41:16&175.585&1.893&0.007&0.374&5\\
FSR 802 &6:01:01&35:16:44&176.167&6.018&0.012&0.613&4\\
FSR 804 &4:34:36&21:41:15&176.505&-17.244&0.005&0.025&4\\
FSR 805 &4:36:27&22:02:32&176.506&-16.685&0.008&0.158&5\\
FSR 809 &5:08:01&27:30:38&176.579&-7.676&0.009&0.439&6\\
FSR 816 &5:39:17&31:30:05&177.099&0.189 &0.007&0.359&4\\
FSR 817 &5:39:27&30:53:36&177.633&-0.104&0.01&0.486&4\\
FSR 823 &5:42:38&29:33:56&179.124&-0.223&0.068&0.137&4\\
FSR 833 &6:05:17&30:47:35&180.541&4.62&0.019&0.153&4\\
FSR 840 &4:37:33&16:29:53&181.238&-19.946&0.036&0.072&4\\
FSR 842 &5:34:22&25:35:44&181.507&-3.89&0.01&0.504&4\\
FSR 846 &5:48:44&26:22:05&182.555&-0.739&0.01&0.486&4\\
FSR 848 &6:34:30&31:23:25&182.893&10.455&0.019&0.077&4\\
FSR 849 &5:51:13&25:46:18&183.352&-0.568&0.007&0.091&4\\
FSR 850 &5:45:15&24:45:13&183.528&-2.249&0.007&0.344&4\\
FSR 861 &5:23:16&18:44:46&185.889&-9.772&0.012&0.072&4\\
FSR 864 &5:47:51&21:55:34&186.26&-3.201&0.009&0.15&5\\
FSR 868 &5:24:56&18:18:21&186.482&-9.681&0.071&0.353&5\\
FSR 888 &6:22:13&23:24:33&188.853&4.437&0.036&0.073&4\\
FSR 890 &6:23:10&23:11:13&189.152&4.527&0.007&0.333&5\\
FSR 893 &6:13:45&21:32:54&189.572&1.833&0.006&0.302&5\\
FSR 907 &5:29:27&13:21:27&191.326&-11.41&0.052&0.103&5\\
FSR 925 &6:05:05&16:06:40&193.34&-2.592&0.013&0.671&4\\ 
FSR 929 &6:25:32&17:43:12&194.26&2.477&0.007&0.333&4\\
FSR 944 &7:21:48&22:29:50&195.653&16.448&0.024&0.071&6\\
FSR 946 &6:10:58&14:09:30&195.74&-2.293&0.005&0.02&4\\
FSR 947 &6:08:59&13:52:34&195.754&-2.853&0.01&0.52&4\\
FSR 957 &6:25:22&14:34:59&197.017&0.977&0.007&0.36&4\\
FSR 963 &6:14:51&12:51:31&197.333&-2.088&0.022&0.089&5\\
FSR 964 &5:24:26&5:28:12&197.626&-16.548&0.006&0.292&4\\
FSR 966 &5:17:30&4:24:29&197.668&-18.573&0.007&0.323&6\\
FSR 967 &6:29:19&14:14:08&197.77&1.663&0.034&0.102&5\\
FSR 968 &6:11:21&11:51:39&197.802&-3.315&0.003&0.022&5\\
\hline
\end{tabular}
\begin{list}{Table Notes.}
\item Cols. $2-3$: Central coordinates provided by FSR07. Cols. $4-5$: Corresponding Galactic coordinates. Cols. $6-7$: Core and tidal radii derived by FSR07 from King fits. Col. $8$: FSR quality flag.
\end{list}
}
\end{table}

\begin{figure*}
\begin{center}
\psfragscanon
\psfrag{a}[b][b][0.6][0]{$\alpha(2000)$}
\psfrag{b}[b][b][0.6][0]{$\delta(2000)$}
\psfrag{c}[b][b][0.6][0]{$\alpha(2000)$}
\psfrag{        d}[b][b][0.6][0]{$\delta(2000)$}
   \includegraphics[scale=0.175,angle=-90,viewport=0 0 920 920,clip]{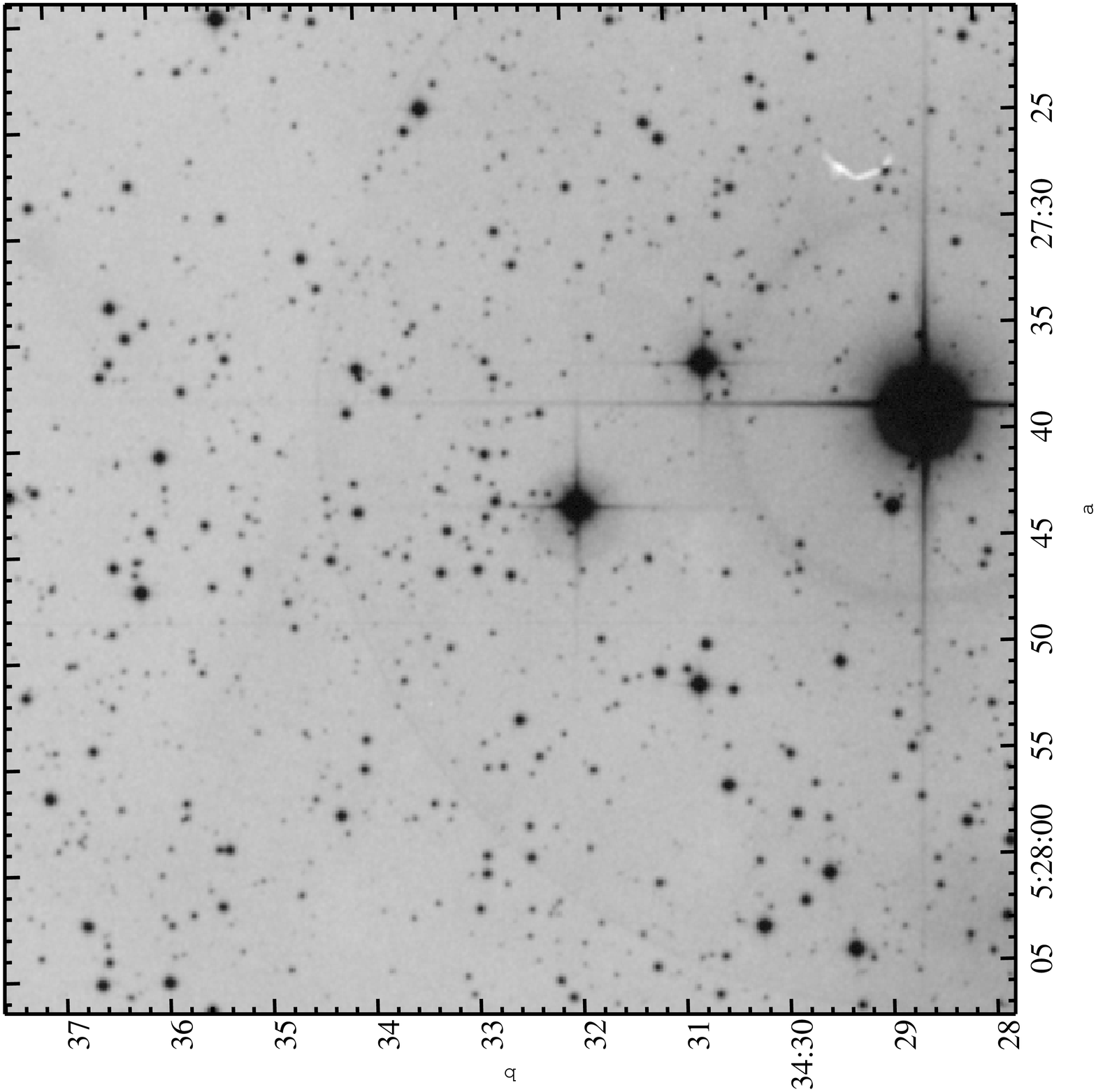}
   \includegraphics[scale=0.18,angle=-90,viewport=0 0 920 920,clip]{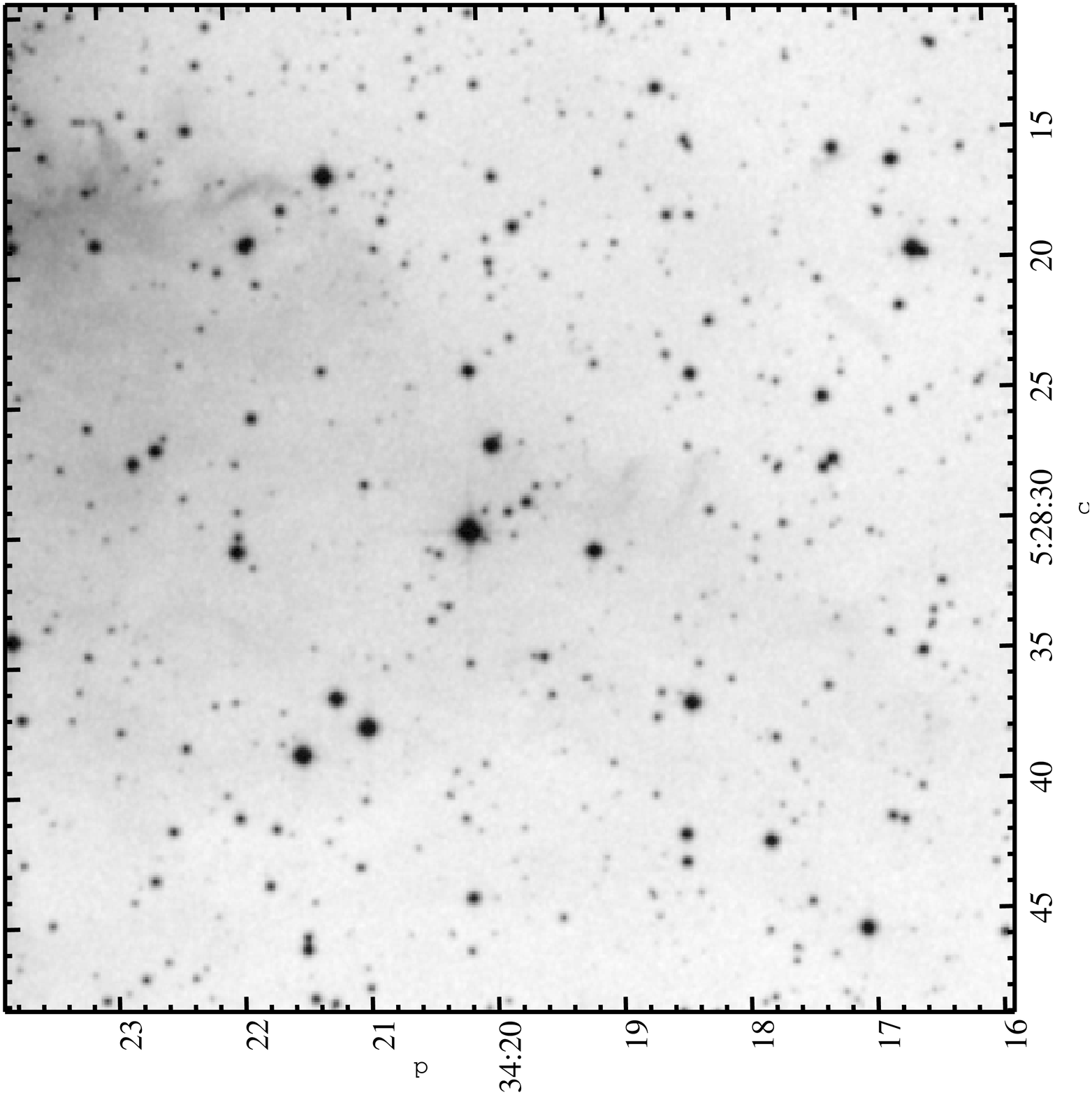}
   \includegraphics[scale=0.18,angle=-90,viewport=0 0 920 920,clip]{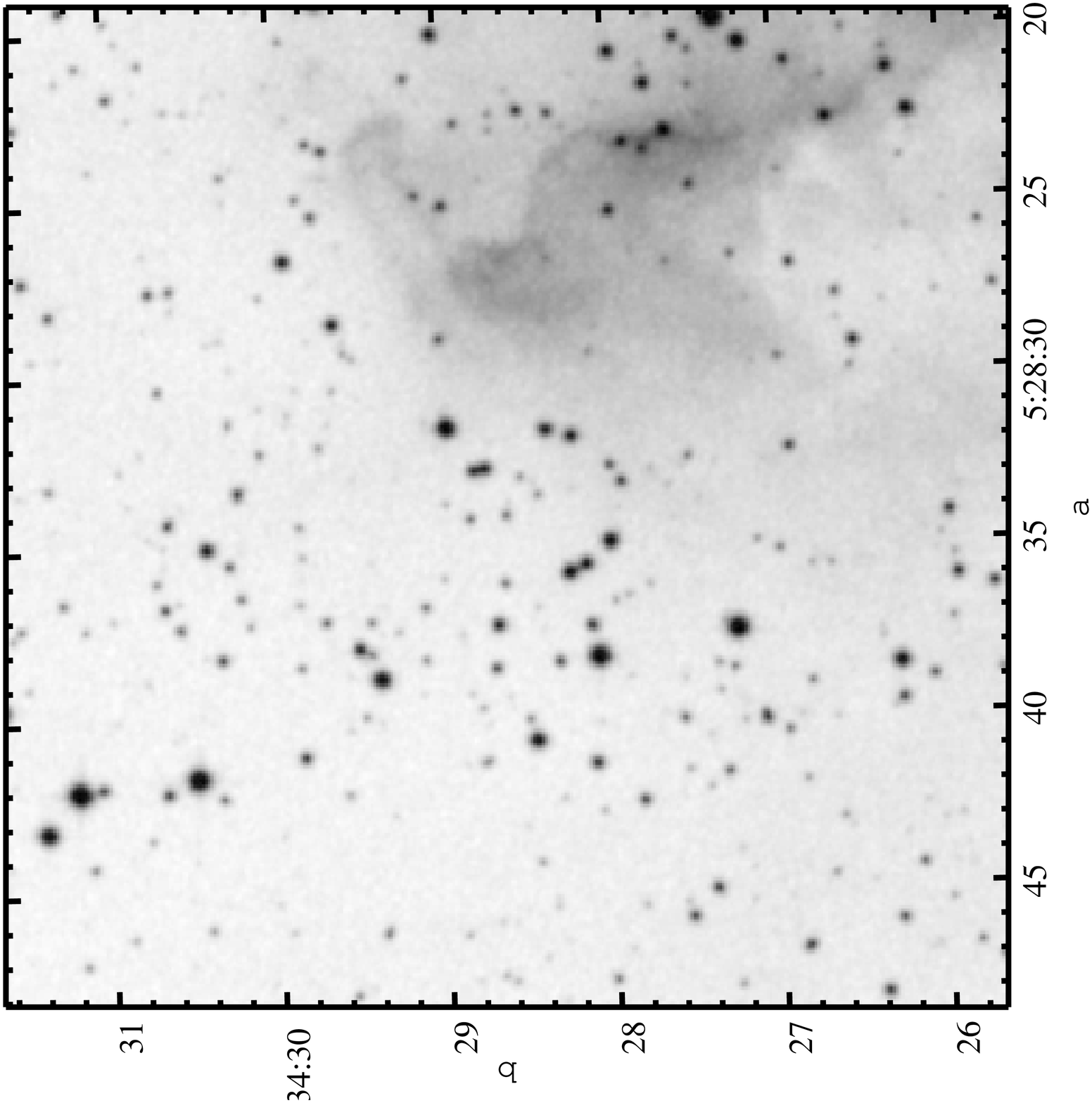}
   \includegraphics[scale=0.18,angle=-90,viewport=0 0 920 920,clip]{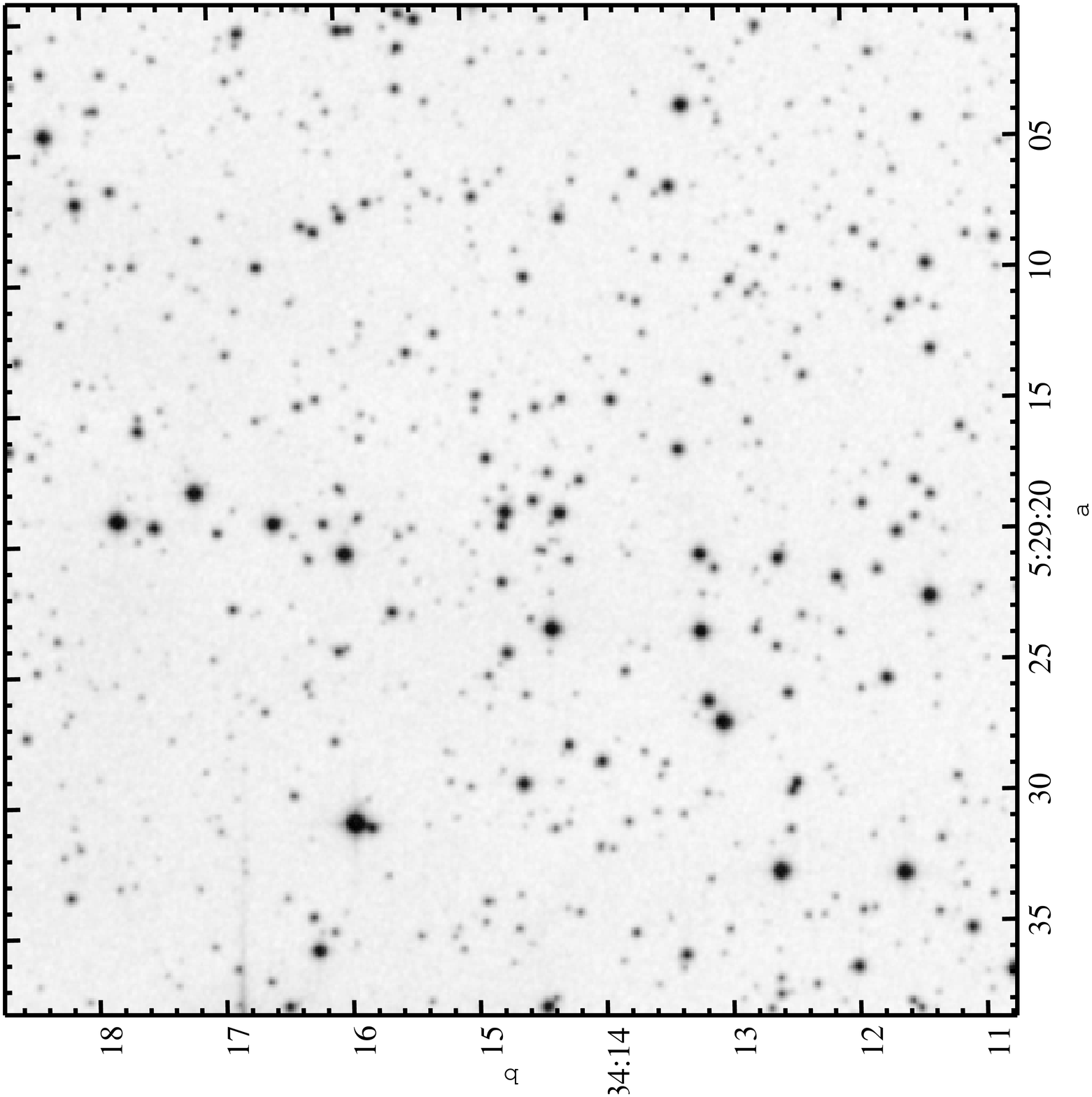}
   \includegraphics[scale=0.18,angle=-90,viewport=0 0 920 920,clip]{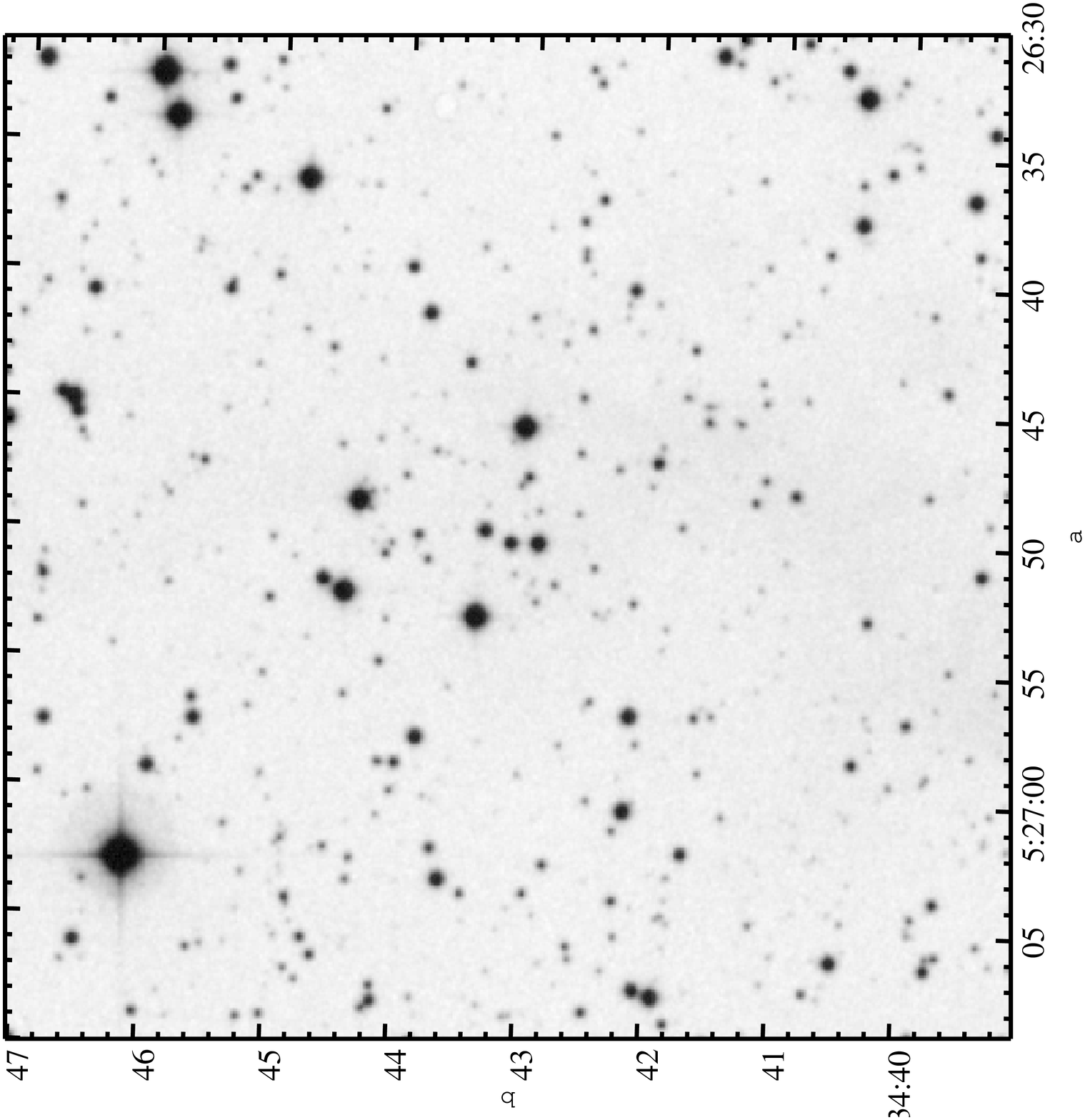}
   \includegraphics[scale=0.175,angle=-90,viewport=0 0 920 920,clip]{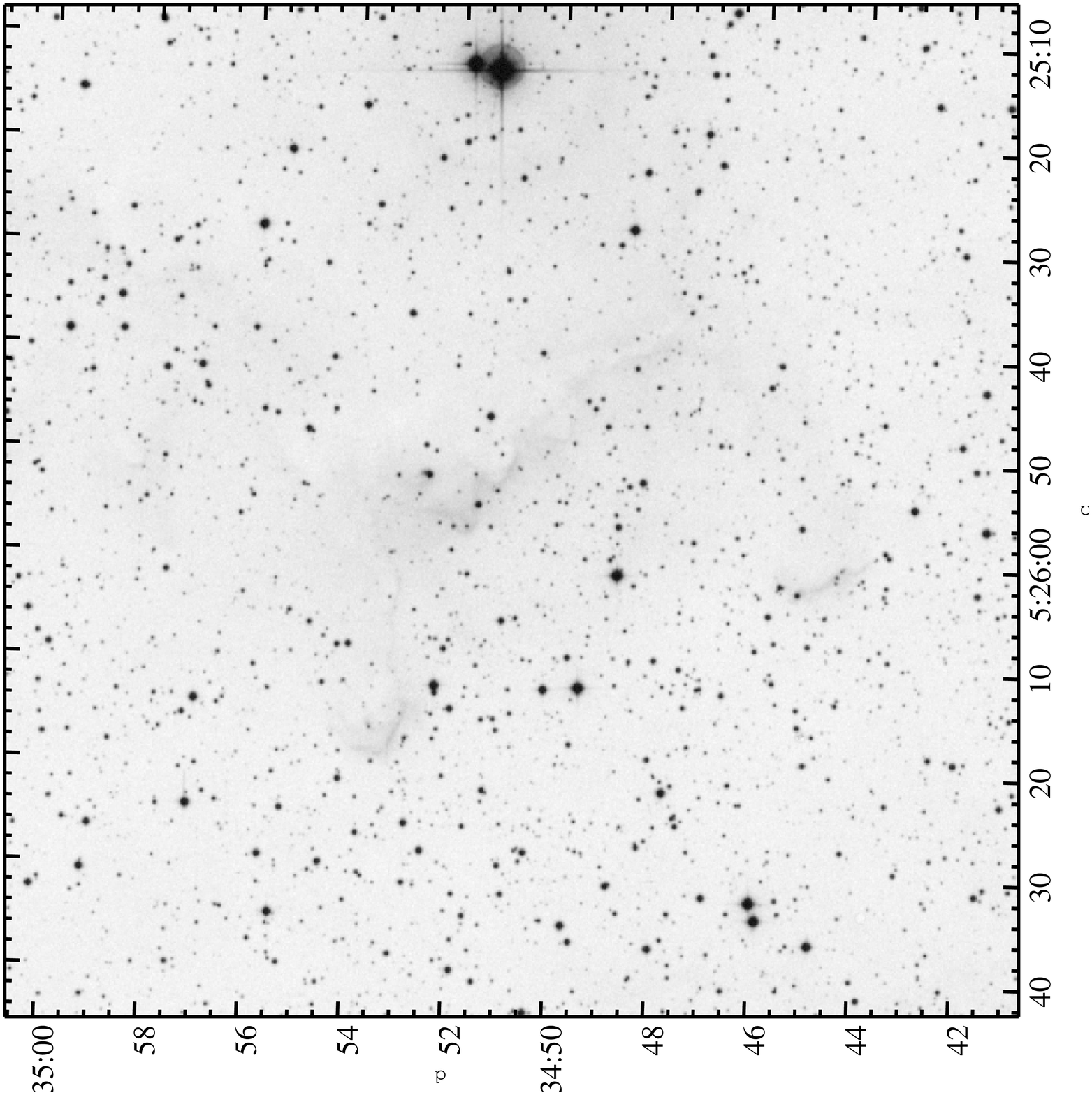}
   \includegraphics[scale=0.18,angle=-90,viewport=0 0 920 920,clip]{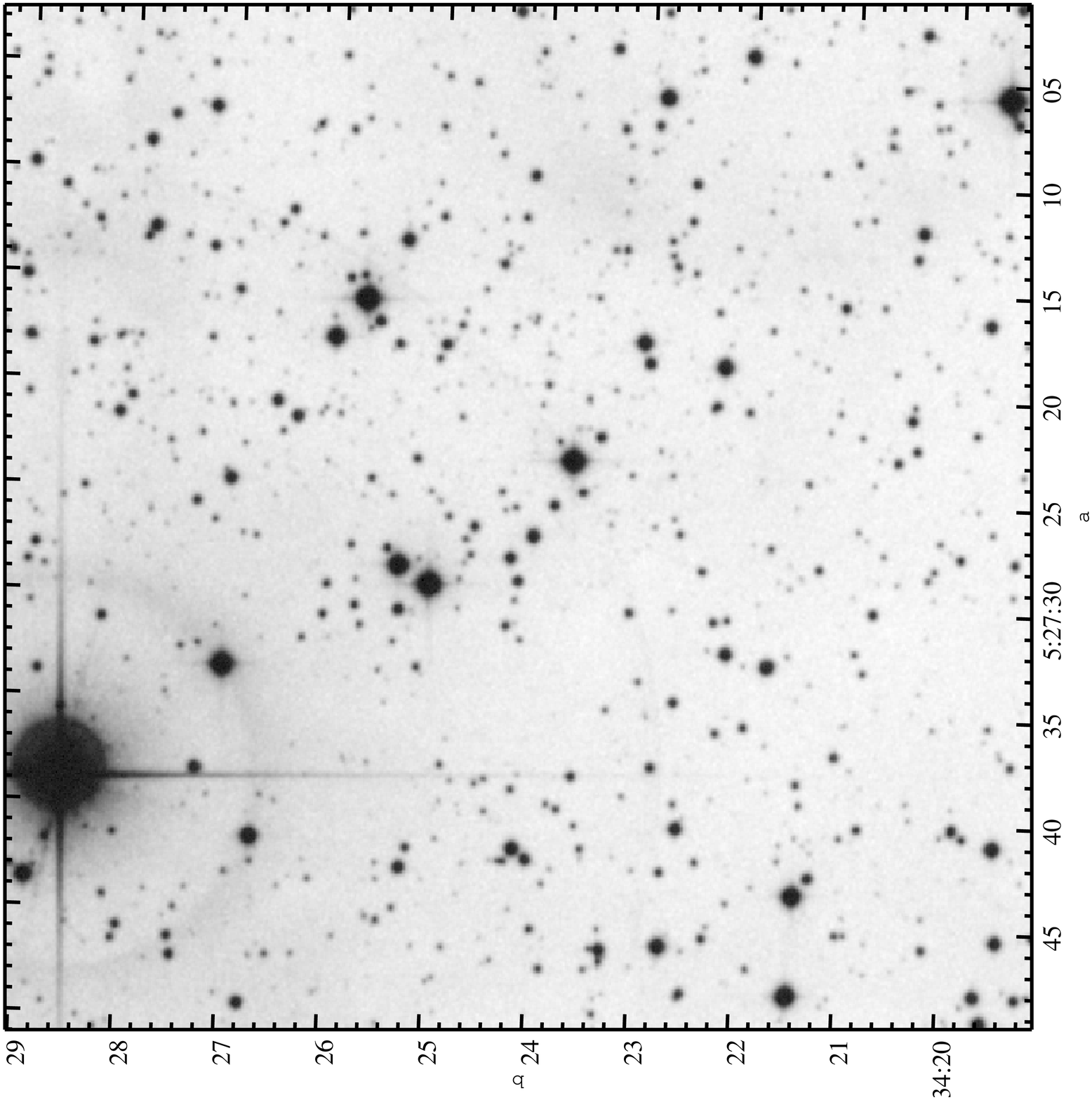}
   \includegraphics[scale=0.18,angle=-90,viewport=0 0 920 920,clip]{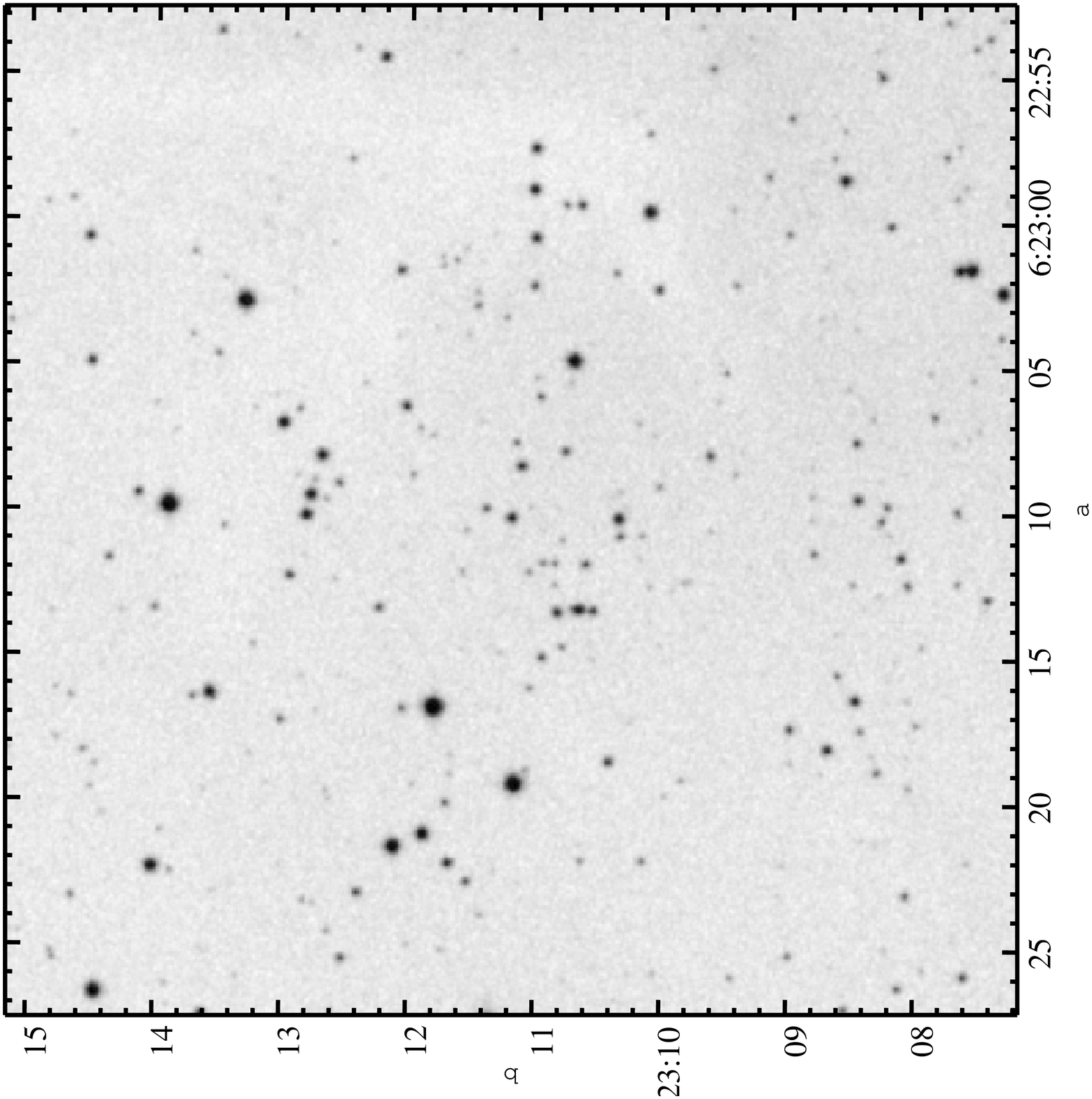}
   \includegraphics[scale=0.18,angle=-90,viewport=0 0 920 920,clip]{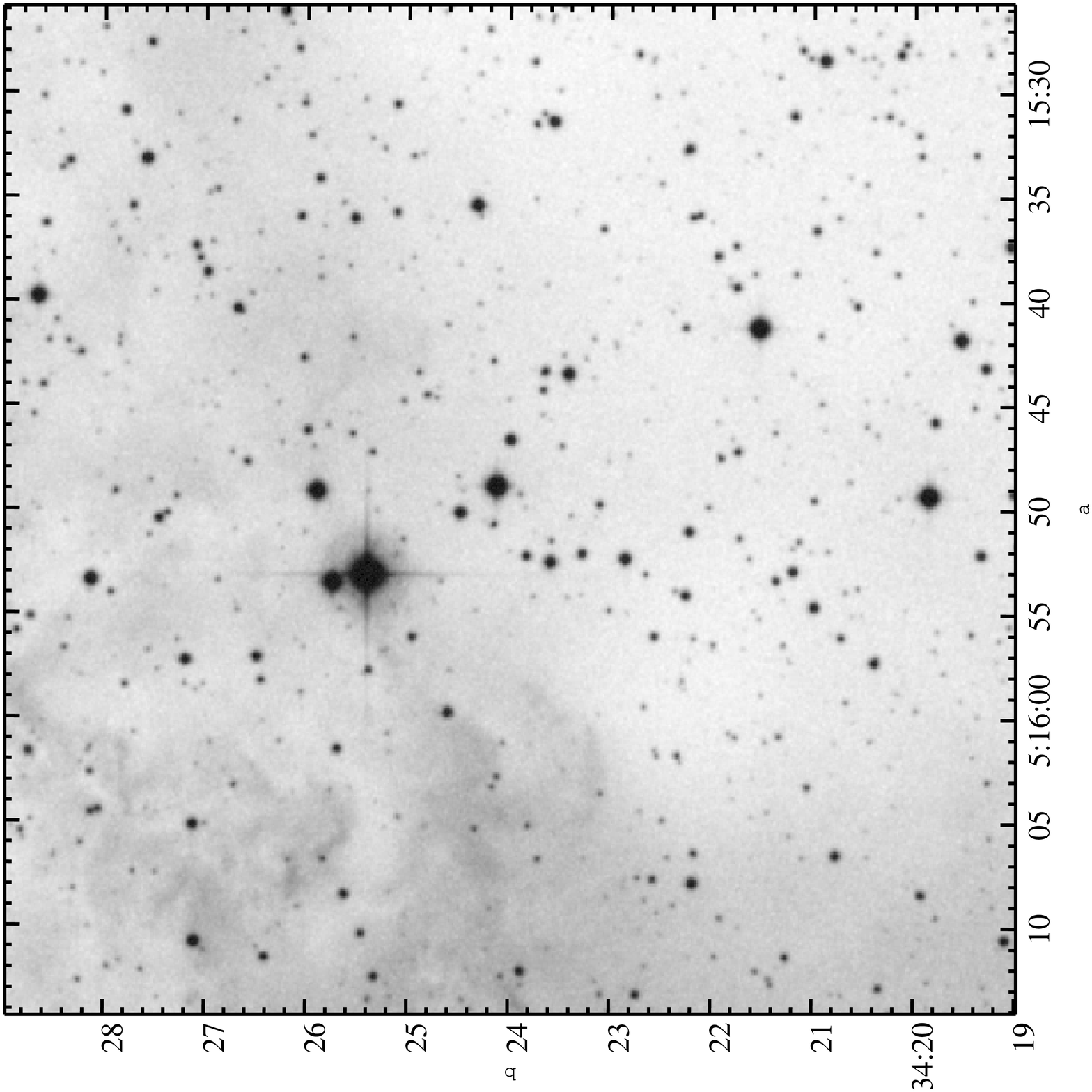}
   \caption[]{First row panels: XDSS R image centred on CBB 3 ($10'\times10'$), CBB 4 ($8'\times8'$), and CBB 5 ($6'\times6'$). Second row: the same for CBB 6 ($8'\times8'$), CBB 7 ($8'\times8'$), and CBB 9 ($15'\times15'$). Third row: FSR $780$ ($10'\times10'$), FSR $890$ ($8'\times8'$), and CBB 8 ($10'\times10'$).}
\end{center}
   \label{group}
\end{figure*}

Most young star clusters dissolve in the Galactic field in the early phase of their existence because of the rapid primordial gas removal by winds from OB stars and supernova explosions (\textit{infant mortality}), since the potential of an embedded cluster (EC) is generally dominated by gas \citep{Tutukov78, Goodwin06, Bonatto11b}. It is accepted that at this stage, the fate of a cluster is determined by the star formation efficiency (SFE) and the mass of the more massive stars. If the EC blows out the gas adiabatically, the cluster will remain bound as long as the SFE is higher than $30\%$, but if the gas expulsion is explosive, the SFE needs to be higher than $50\%$ \citep{Lada03}. Therefore, the gas expulsion can be very disruptive and because of this $\approx\,95\%$ of the ECs do not survive to become OCs \citep{Lada03, Bonatto11b}, with the survivors keeping at most $50\%$ of their stars \citep[\textit{infant weight loss} - ][]{Kroupa02, Weidner07, Goddard10}. On the other hand, \citet{Smith11} argue that the variation in cluster initial conditions is the most important parameter for dissolution by gas expulsion. This occurs because the stellar distribution can change significantly, changing the relative importance of the stellar and gas potentials. 

Star formation occurs inside massive and dense gas clumps in giant molecular clouds (GMCs). These structures contain many cores that form stars. Some cores can group themselves in small sub-clumps where the SFE can be higher than in the overall clump. This way, \citet{Goodwin09} suggest that the determinant factor for cluster survival is the virial state of the stellar content immediately before the gas expulsion. In this context, both \textit{infant mortality} and cluster infant stellar loss depend on the radial density profile (RDP) just before the gas expulsion \citep{Boily03} and the relative distribution of stars and gas \citep{Adams00}.

\begin{table}
{\footnotesize
\caption{Cross-identification of the open clusters.}
\renewcommand{\tabcolsep}{2.3mm}
\renewcommand{\arraystretch}{1.1}
\begin{tabular}{lrrr}
\hline
\hline
Designations&References\\
\hline
FSR 716,\,SAI\,44&3, 7\\
FSR 784,\,Koposov\,7,\,Sh2-235 North-West&3, 4, 5, 7, 8\\
FSR 849,\,Koposov\,58&3, 7\\
ASCC\,23,\,FSR 746&1, 2, 3\\
FSR 802,\,Koposov\,12&3, 4, 7\\
Luginbuhl-Skiff\,1,\,Skiff J0614+12.9,\,FSR 963&1, 3, 4, 6\\
\hline
\end{tabular}
\begin{list}{Table Notes.}
\item The references are: 1 - \citet{Dias02}; 2 - \citet{Kharchenko05b}; 3 -  FSR07; 4 - \citet{Koposov08}; 5 - \citet{Kirsanova08}; 6 - \citet{Tadross08}; 7 - \citet{Glushkova10}; 8 - \citet{Camargo11}.
\end{list}
\label{tab2}
}
\end{table}

Observations indicate that the majority, if not all, of the star formation takes place in clusters \citep{Lada03, Allen07} thus, these objects are important tracers of the stellar population properties in the Galaxy. In addition, there is a connection between star cluster dissolution (or stellar mass loss) and the field star population \citep{Massey95, Chandar06}. Given the importance of the OCs for our understanding of the Galaxy, many catalogues and sky surveys were compiled, especially in recent years \citep{Alter70, Lynga87, Dias02, Dutra03, Bica03a, Bica03b, Mermilliod03, Bica05, Kharchenko05a, Kharchenko05b, Froebrich07, Koposov08, Glushkova10}]. 

\begin{figure}
\resizebox{\hsize}{!}{\includegraphics{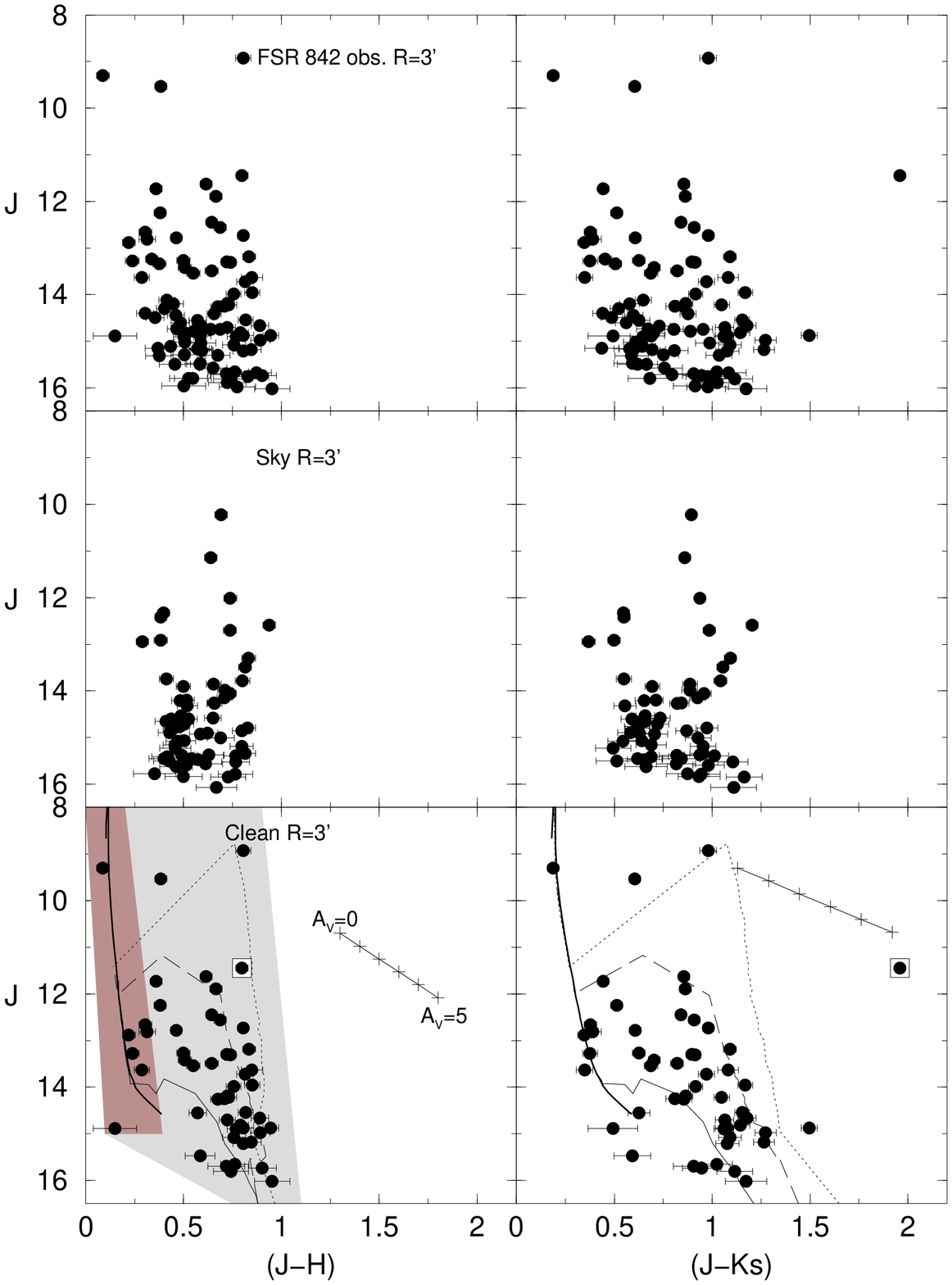}}
\caption[]{2MASS CMDs extracted from the central region of FSR 842. Top panels: observed CMDs $J\times(J-H)$ (left) and $J\times(J-K_s)$ (right). Middle panels: equal area comparison field. Bottom panels: field star decontaminated CMDs fitted with the 5 Myr MS Padova isochrone (solid line) and PMS isochrones of Siess, 0.1 (dotted line), 1 (dashed line), and 10Myr (solid line). The colour-magnitude filters used to isolate cluster MS/evolved and PMS stars are shown as shaded regions. We also present the reddening vector for $A_V=0$ to 5. The square indicates a B star.}
\label{cmd2}
\end{figure}

\begin{figure}
\resizebox{\hsize}{!}{\includegraphics{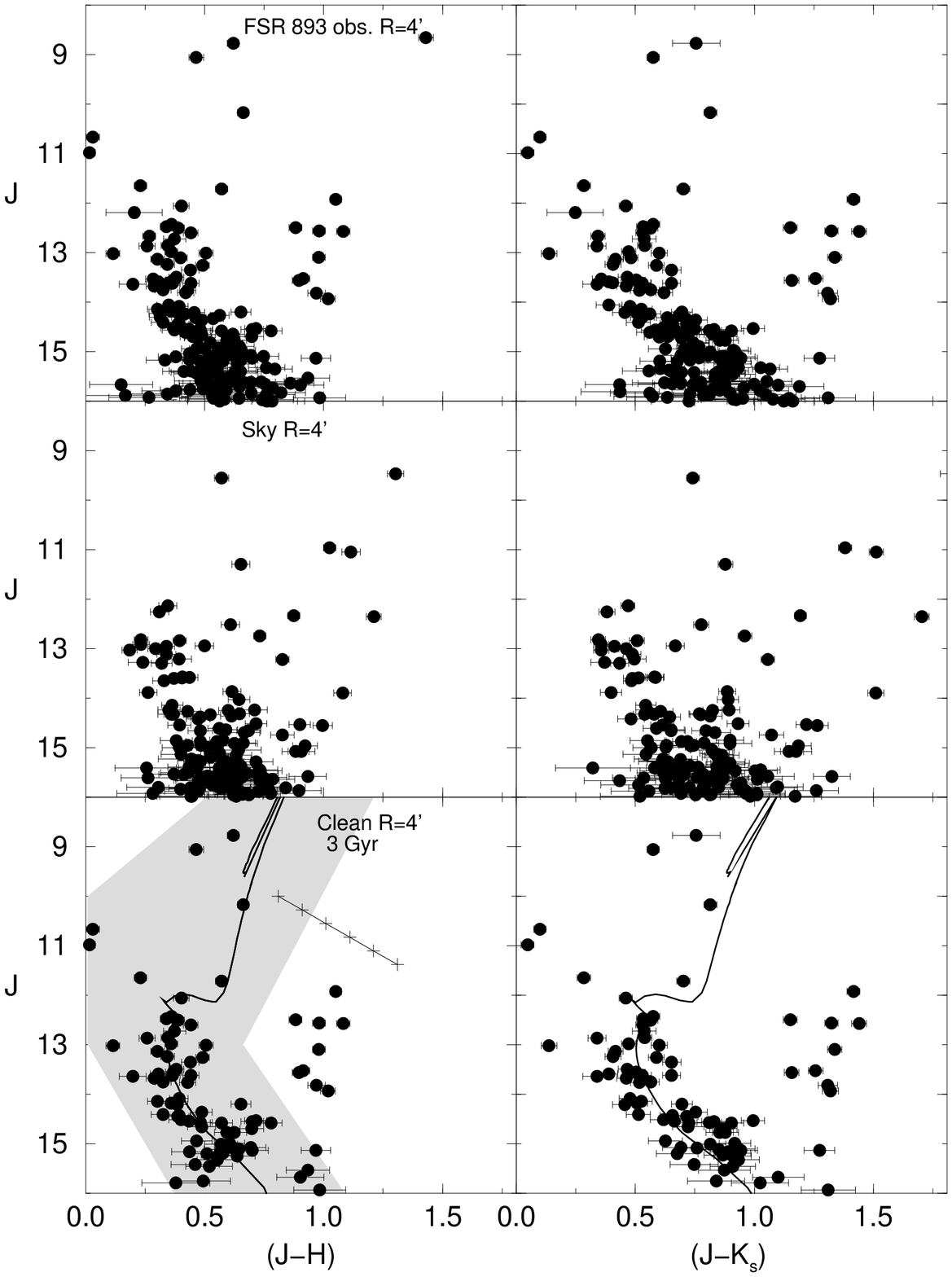}}
\caption[]{2MASS CMDs extracted from the $R=4'$ region of FSR 893. Top panels: observed CMDs $J\times(J-H)$ (left) and $J\times(J-K_s)$ (right). Middle panels: equal area comparison field. Bottom panels: field-star decontaminated CMDs fitted with the 3 Gyr Padova isochrone (solid line). The colour-magnitude filter used to isolate cluster MS/evolved stars is shown as a shaded region.}
\label{FSR893}
\end{figure}

OCs do not appear to form in isolation, but in associations of clusters embedded in the same star complex \citep{Efremov78}. Generally, these structures are linked to Galactic spiral arm systems, which is consistent with the fact that spiral arm encounters are efficient generators of GMC instability, leading to fragmentation and collapse. Star formation followed by supernova explosions, stellar winds from massive stars, and H II region expansions can disrupt a GMC on a timescale of a few $10^7$yr \citep{Elmegreen00, Bonnell06}, populating the Galactic disk with dynamical families of clusters, i.e., groups of OCs that have a common dynamical origin \citep{King03, Fuente08}. These families include 10-20 objects and disperse on a timescale of $\sim20$ Myr. Later, the star complex evolves to individual clusters and the families cannot be recognised any more, since most clusters have been disrupted.
   
On the other hand, a GMC may fragment into some small nebulae that are scenarios of star formation on a smaller scale. A nebula may collapse to form some clusters close to each other, because of the action of massive stars that may trigger sequential star formation. These systems differ from families of clusters that are formed in a star complex with a scale of about 600 pc.

The fate of these objects depends on several factors, but if the structure survives the \textit{infant mortality} as a stable bound system, they have a good chance to form cluster pairs or multiple systems (a cluster of clusters, \citealp{Bica99} in the LMC and  \citealp{Feigelson11} in the Galaxy). However, if the group forms a bound system, but does not reach equilibrium, they may merge to form massive clusters. On the other hand, if after the phase of primordial gas removal the surviving objects of this structure form an unbound system, they evolve to an association of clusters and eventually disperse. The fact that multiple systems are rare after the gas expulsion, suggests that they are extremely unstable with merging or tidal disruption timescales of a few Myr, probably lesser than the age spread of stars inside an OC\footnote{\citet{Fuente09} estimated a timescale $<16$ Myr for cluster pairs in-contact and according to \citet{Fellhauer09} the merging time-scale for sub-clumps, is shorter than the gas removal time.}. However, there are evidence that associations of clusters and sub-structured clusters may develop high local SFEs reducing the effect of the gas expulsion and favouring bound cluster formation \citep{Goodwin09, Moeckel11, Kruijssen12}. It is probable that some young clusters presenting substructures and RDPs that do not follow King's law result for a group of merging clusters, and the age spread may indicate the duration of star formation in the primordial association of OCs. 

The main goal of this work is to improve the census of the star clusters towards the Galactic anticentre and derive their basic parameters.  
We investigate the nature of 48 stellar overdensities from the catalogue of \citet[][hereafter FSR07]{Froebrich07} located in the sector $160^\circ\,\leq\,\ell\,\leq 200^\circ$, labelled by them with quality flags 4, 5, and 6 (Table \ref{tab1}). We also derive parameters for Stock 8, Kronberger 1 (DSH J0528.3+3446), BPI 14 and for 7 clusters discovered in the present work (CBB 3 to CBB 9), resulting in 58 analysed objects.

\begin{figure}
\resizebox{\hsize}{!}{\includegraphics{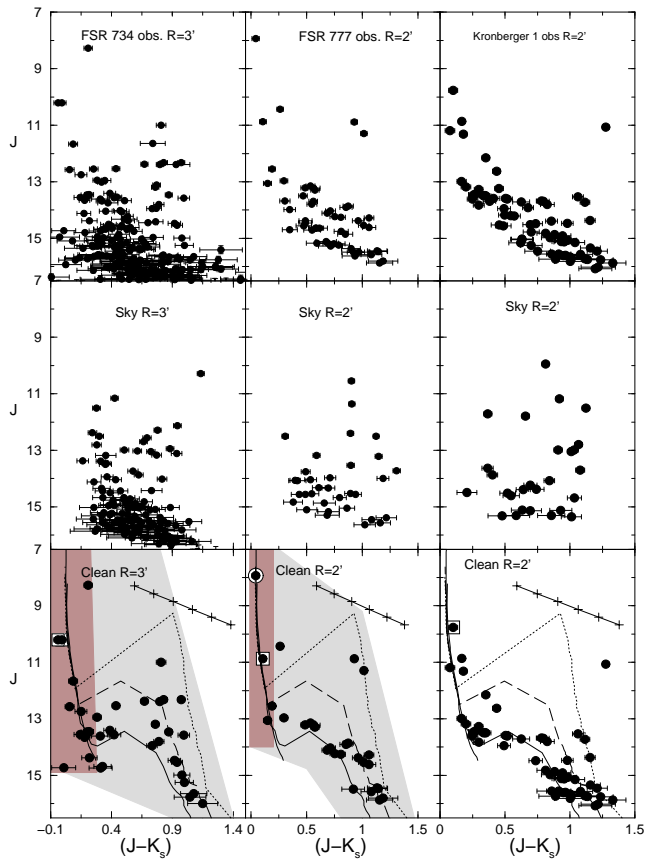}}
\caption[]{2MASS CMDs for the clusters FSR 734, FSR 777, and Kronberger 1. Top panels: observed CMDs $J\times(J-K_s)$. Middle panels: equal area comparison field. Bottom panels: field star decontaminated CMDs fitted with the 2 Myr MS Padova isochrone (solid line) for FSR 734 and 3 Myr for the others, and PMS isochrones of Siess, 0.1 (dotted line), 1 (dashed line), and 5 Myr (solid line). The colour-magnitude filters used to isolate cluster MS/evolved and PMS stars are shown as shaded regions. We also present the reddening vector for $A_V=0$ to 5.}
\label{cmd1}
\end{figure}

\begin{figure}
\resizebox{\hsize}{!}{\includegraphics{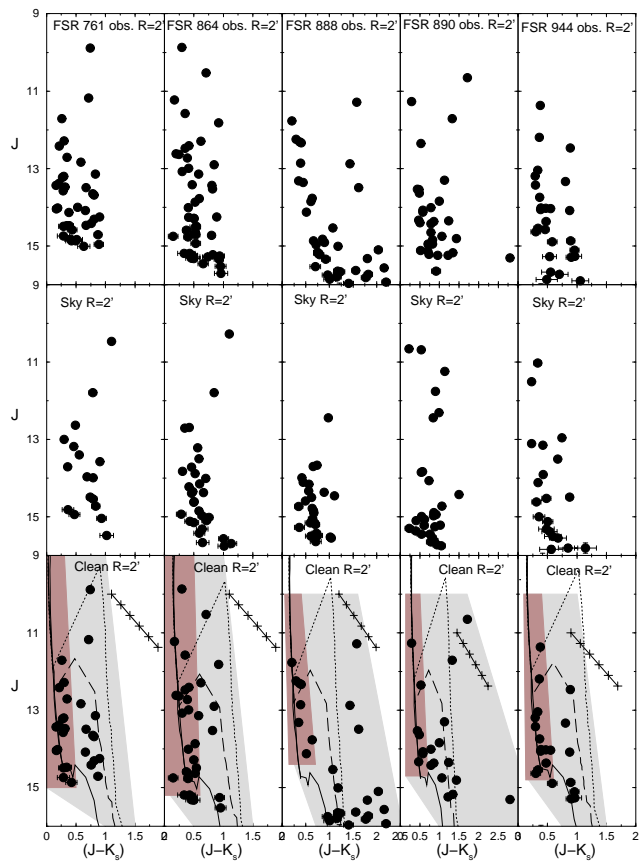}}
\caption[]{Same as Fig.~\ref{cmd1} for the clusters FSR 761, FSR 864, FSR 888, FSR 890, and FSR 944. The MS Padova isochrone used were 2 Myr for FSR 761 and 3 Myr for others, and the PMS isochrones of Siess, 0.1 (dotted line), 1 (dashed line), and 10 Myr (solid line).}
\label{cmd4}
\end{figure}

This paper is organised as follows. In Sect. \ref{sec:2} we list the cluster candidates. In Sect. \ref{sec:3} we present the 2MASS photometry the methods employed in the CMD analyses and derive fundamental parameters (\textit{age, reddening, distance}) for the overdensities shown to be clusters.  Sect. \ref{sec:4} focuses on cluster structural parameters. In Sect. \ref{Mass} we estimate the mass for the clusters with derived structural parameters. In Sect. \ref{sec:7} we discuss the results.  Finally, in Sect. \ref{sec:8} we provide concluding remarks.

\section{Cluster candidates}
\label{sec:2}

FSR07 provided a catalogue of 1021 star cluster candidates identified in the 2MASS database with $|b|\,\leq\,20^\circ$ along the Galactic plane. They classified the overdensities in probable and possible clusters. Eighty-seven probable clusters and 90 possible clusters candidates are distributed towards the Galactic anticentre. \citet{Bonatto08} analysed 28 FSR cluster candidates projected nearly towards the anti-centre ($160^\circ\,\leq\,\ell\,\leq 200^\circ$) and confirmed 6 new and 9 previously known OCs, 6 uncertain cases and 7 probable fluctuations of the stellar field. \citet{Camargo10} analysed 50 overdensities in the same region, classified as OC candidates with quality flags 2 and 3, finding 16 OCs and 5 uncertain cases.

Table \ref{tab1} lists the present FSR overdensity sample and Table \ref{tab2} shows identifications for those previously studied. The objects selected for the present work are classified by FSR07 as probable and possible OCs and labelled with quality flags 4, 5, and 6.

In Fig.~\ref{group} we illustrate XDSS images in the R band of FSR 780 and FSR 890 as examples of overdensities from FSR07 confirmed as clusters, and of the clusters discovered in the present work.

\begin{figure}
\resizebox{\hsize}{!}{\includegraphics{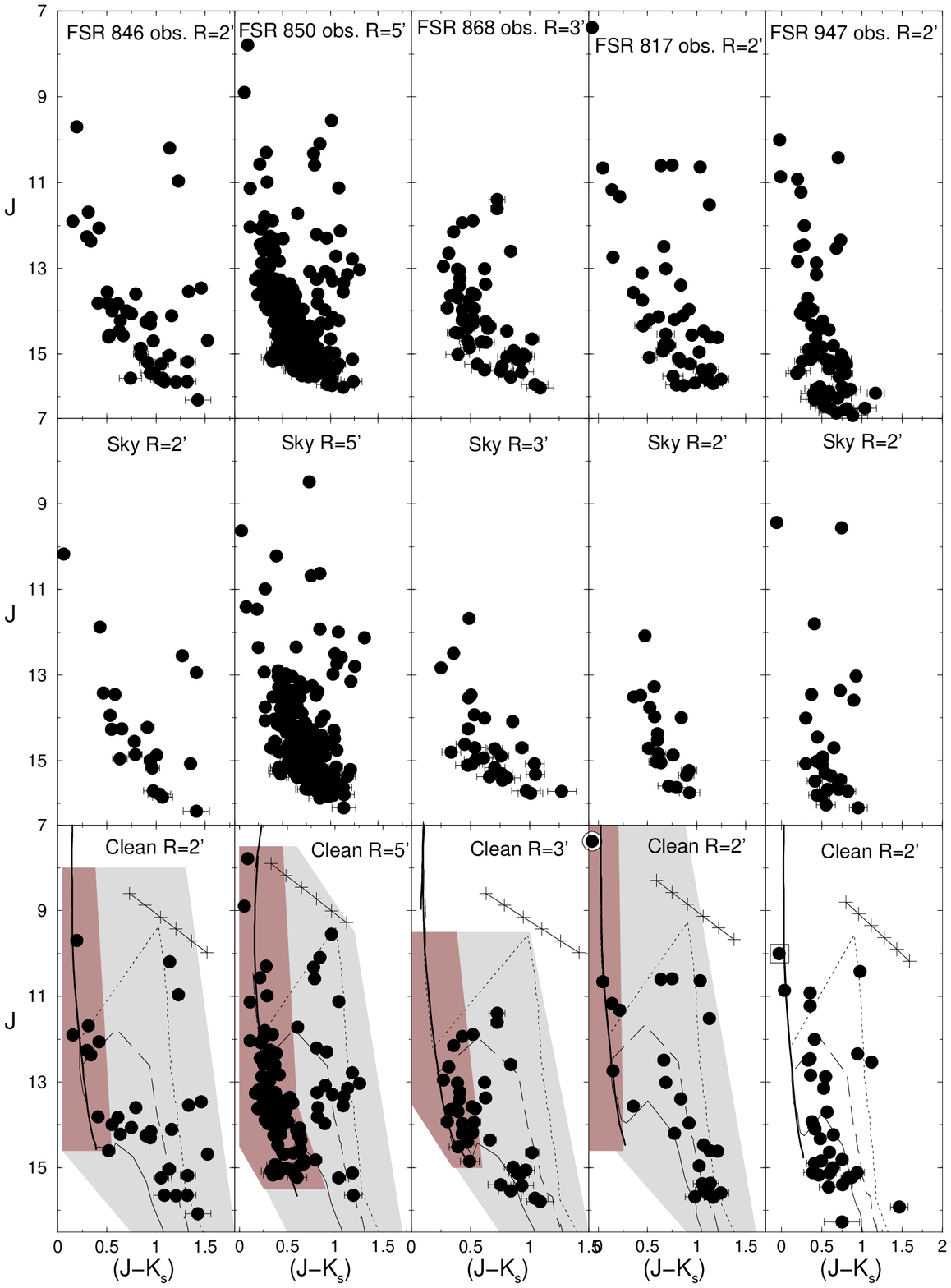}}
\caption[]{Same as Fig.~\ref{cmd1} for the clusters FSR 846, FSR 850, FSR 868, FSR 817, and FSR 947. The PMS were fitted with isochrones of Siess, 0.1 (dotted line), 1 (dashed line), and 5 Myr (solid line).}
\label{cmd5}
\end{figure}

\begin{figure}
\resizebox{\hsize}{!}{\includegraphics{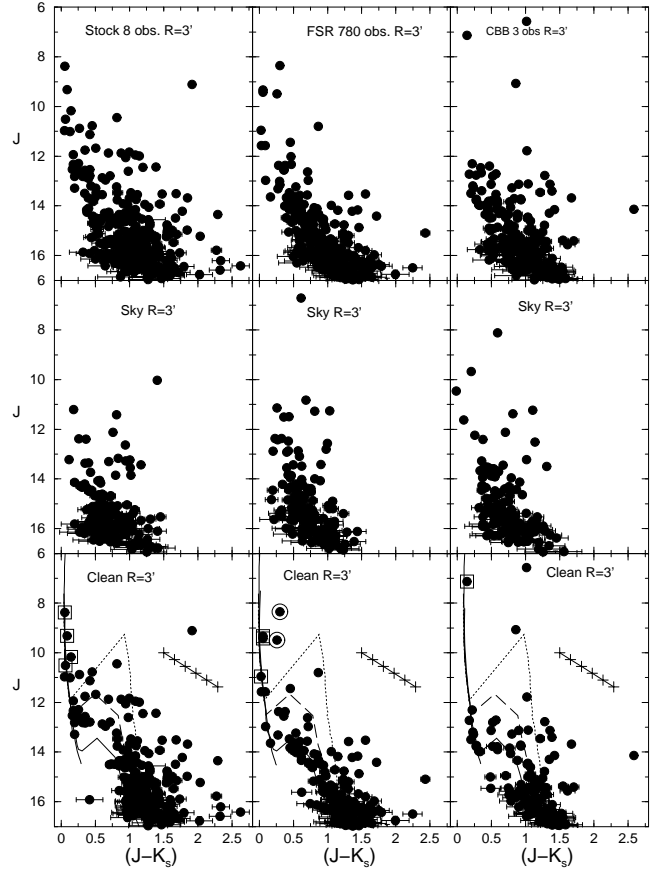}}
\caption[]{Same as Fig.~\ref{cmd1} for the clusters Stock 8, FSR 780, and CBB 3.}
\label{CBB5}
\end{figure}

\section{2MASS photometry}
\label{sec:3}

We use 2MASS\footnote{The Two Micron All Sky Survey, available at \textit{www..ipac.caltech.edu/2mass/releases/allsky/}} photometry \citep{Skrutskie06} in the $J$, $H$ and $K_{s}$ bands extracted in circular concentric regions centred on the coordinates of the OC candidates (Tables~\ref{tab1} and \ref{tab4}) using VizieR\footnote{http://vizier.u-strasbg.fr/viz-bin/VizieR?-source=II/246.}. Large extraction areas are essential to build the RDPs (Sect.~\ref{sec:4}) with a high contrast relative to the background, and for a consistent field star decontamination (Sect.~\ref{sec:3.1}). In addition, 2MASS provides the spatial and photometric uniformity required for relatively high star count statistics.

\subsection{Field-star decontamination}
\label{sec:3.1}

The field-star decontamination algorithm developed by \citet{Bonatto07a} was applied to uncover the intrinsic CMD morphology from the background stars. The use of field-star decontamination to build CMDs
has proved to constrain age and distance more clearly than the observed photometry, especially for low-latitude OCs \citep[e.g.][and references therein]{Bica08}.

The decontamination algorithm is described in detail in \citet{Bonatto07b} and \citet{Bica08}. Here we provide a brief description. The algorithm divides the CMD into a 3D cell grid with axes along the $J$ magnitude and the $(J-H)$  and $(J-K_s)$  colours. It computes the expected number-density of field stars in each cell based on the number of comparison field stars (within the $1\sigma$ Poisson fluctuation) with magnitude and colours compatible with those of the cell. It subtracts the expected number of field stars from each cell. Typical cell dimensions are $\Delta{J}=1.0$, and  $\Delta(J-H)={\Delta(J-K_{s})}=0.2$, which are large enough to allow sufficient statistics in individual cells and small enough to maintain the CMD evolutionary sequences.

\subsection{Fundamental parameters}
\label{sec:3.2}

Fundamental parameters are derived with solar-metallicity Padova isochrones \citep{Girardi02} and \citet{Siess00}, for main sequence (MS) and pre-main sequence stars (PMS), respectively. We estimate the fundamental parameters by \textit{eye}, using the decontaminated CMD morphology. Parameters derived are the observed distance modulus $(m-M)_{J}$ and reddening $E(J-H)$, which convert to $E(B-V)$ and $A_{V}$ with the relations $A_{J}/{A_{V}}=0.276$, $A_{H}/{A_{V}}=0.176$, $A_{K_{s}}/{A_{V}}=0.118$, $A_{J}=2.76\times{E(J-H)}$  and $E(J-H)=0.33\times{E(B-V)}$ \citep{Dutra02}, assuming a constant total-to-selective absorption ratio $R_{V}=3.1$ \citep{Cardelli89}. 

\begin{table*}
{\footnotesize
\begin{center}
\caption{Derived fundamental parameters for 28 confirmed clusters.}
\renewcommand{\tabcolsep}{1.4mm}
\renewcommand{\arraystretch}{1.3}
\begin{tabular}{lrrrrrrrrr}
\hline
\hline
Cluster&$\alpha(2000)$&$\delta(2000)$&$\aV$&Age&$d_{\odot}$&$R_{GC}$&$x_{GC}$&$y_{GC}$&$z_{GC}$\\
&(h\,m\,s)&$(^{\circ}\,^{\prime}\,^{\prime\prime})$&(mag)&(Myr)&(kpc)&(kpc)&(kpc)&(kpc)&(kpc)\\
($1$)&($2$)&($3$)&($4$)&($5$)&($6$)&($7$)&($8$)&($9$)&($10$)\\
\hline
BPI 14&05:29:00&34:24:00&$1.98\pm0.20$&$1\pm1$&$2.7\pm0.3$&$9.90\pm0.3$&$-09.89\pm0.3$&$+0.30\pm0.03$&$-0.01\pm0.01$\\
CBB 3 &05:27:43.31&34:32:36.0&$1.98\pm0.20$&$2\pm1$&$2.7\pm0.3$&$9.90\pm0.3$&$-09.89\pm0.3$&$+0.32\pm0.03$&$-0.01\pm0.01$\\
CBB 4 &05:28:29.3&34:19:50&$1.98\pm0.20$&$2\pm1$&$2.7\pm0.3$&$9.90\pm0.3$&$-09.89\pm0.3$&$+0.30\pm0.03$&$-0.01\pm0.01$\\
CBB 5 &05:28:33.9&34:28:37&$1.98\pm0.20$&$2\pm1$&$2.7\pm0.3$&$9.90\pm0.3$&$-09.89\pm0.3$&$+0.31\pm0.03$&$-0.01\pm0.01$\\
CBB 6 &05:29:19&34:14:41.4&$2.98\pm0.20$&$2\pm1$&$2.7\pm0.5$&$09.93\pm0.5$&$-09.92\pm0.5$&$+0.30\pm0.03$&$0.0\pm0.01$\\
CBB 7 &05:26:50&34:43:10&$2.98\pm0.20$&$2\pm1$&$2.5\pm0.5$&$09.69\pm0.5$&$-09.68\pm0.5$&$+0.30\pm0.03$&$-0.01\pm0.01$\\
CBB 8 &05:15:50&34:24:00&$3.57\pm0.20$&$2\pm1$&$2.41\pm0.7$&$9.61\pm0.7$&$-09.60\pm0.7$&$+0.34\pm0.03$&$-0.10\pm0.01$\\
CBB 9 &05:25:55&34:50:54&$3.27\pm0.20$&$2\pm1$&$2.6\pm0.5$&$9.82\pm0.5$&$-09.82\pm0.5$&$+0.33\pm0.03$&$-0.01\pm0.01$\\
FSR 734 &05:03:22.6&42:25:15.2&$2.18\pm0.20$&$2\pm1$&$2.62\pm0.3$&$09.77\pm0.3$&$-09.74\pm0.3$&$+0.72\pm0.07$&$+0.02\pm0.01$\\
FSR 761 &05:33:23&39:50:44&$2.78\pm0.20$&$2\pm1$&$2.54\pm0.3$&$09.73\pm0.3$&$-09.72\pm0.3$&$+0.47\pm0.04$&$+0.16\pm0.02$\\
FSR 777 &05:27:31&34:44:01&$1.98\pm0.20$&$3\pm2$&$2.69\pm0.3$&$09.89\pm0.3$&$-09.89\pm0.3$&$+0.33\pm0.03$&$-0.01\pm0.01$\\
FSR 780 &05:27:26&34:24:12&$1.98\pm0.20$&$2\pm1$&$2.69\pm0.3$&$9.90\pm0.3$&$-09.89\pm0.3$&$+0.31\pm0.03$&$-0.01\pm0.01$\\
FSR 816 &5:39:17&31:30:05&$1.98\pm0.20$&$10\pm5$&$1.78\pm0.5$&$8.99\pm0.5$&$-08.99\pm0.5$&$+0.10\pm0.01$&$0.01\pm0.01$\\
FSR 817 &5:39:27&30:53:36&$1.98\pm0.20$&$2\pm2$&$2.3\pm0.3$&$9.56\pm0.3$&$-09.56\pm0.3$&$+0.10\pm0.01$&$0.0\pm0.01$\\
FSR 833 &06:05:15&30:47:55&$1.79\pm0.20$&$3\pm2$&$2.89\pm0.4$&$10.10\pm0.4$&$-10.10\pm0.4$&$-0.03\pm0.01$&$+0.23\pm0.02$\\
FSR 842 &05:34:18.8&25:36:38&$2.68\pm0.2$&$5\pm3$&$1.95\pm0.2$&$09.17\pm0.2$&$-09.17\pm0.2$&$-0.05\pm0.01$&$-0.13\pm0.01$\\
FSR 846 &05:48:44&26:22:05&$2.98\pm0.2$&$3\pm2$&$2.48\pm0.3$&$9.70\pm0.3$&$-9.70\pm0.3$&$-0.11\pm0.01$&$-0.03\pm0.01$\\
FSR 850 &05:45:15&24:45:13&$2.18\pm0.20$&$10\pm5$&$2.75\pm0.5$&$09.96\pm0.5$&$-09.96\pm0.5$&$-0.17\pm0.02$&$-0.11\pm0.01$\\
FSR 864 &05:47:49.9&21:55:32.5&$2.48\pm0.2$&$5\pm3$&$2.90\pm0.3$&$10.10\pm0.3$&$-10.10\pm0.3$&$-0.32\pm0.03$&$-0.16\pm0.02$\\
FSR 868 &05:24:56&18:18:21&$2.98\pm0.20$&$5\pm3$&$2.72\pm0.3$&$09.90\pm0.3$&$-09.88\pm0.3$&$-0.30\pm0.03$&$-0.46\pm0.04$\\
FSR 888 &06:22:13&23:24:33&$3.17\pm0.2$&$3\pm2$&$2.65\pm0.3$&$09.84\pm0.3$&$-09.83\pm0.3$&$-0.41\pm0.04$&$+0.21\pm0.02$\\
FSR 890 &06:23:10&23:11:13&$3.37\pm0.2$&$3\pm2$&$2.58\pm0.3$&$09.77\pm0.3$&$-09.76\pm0.3$&$-0.41\pm0.04$&$+0.20\pm0.02$\\
FSR 893 &06:13:45&21:32:54&$0.99\pm0.06$&$3000\pm1500$&$1.1\pm0.5$&$08.3\pm0.5$&$-08.31\pm0.5$&$-0.18\pm0.02$&$+0.04\pm0.01$\\
FSR 944 &07:21:48&22:29:50&$3.17\pm0.2$&$3\pm2$&$2.42\pm0.3$&$09.5\pm0.3$&$-09.45\pm0.3$&$-0.63\pm0.06$&$+0.68\pm0.07$\\
FSR 946 &06:10:58 &14:09:30&$4.46\pm0.2$&$1\pm1$&$2.05\pm0.3$&$09.21\pm0.3$&$-09.19\pm0.3$&$-0.56\pm0.05$&$-0.08\pm0.01$\\
FSR 947 &06:08:59&13:52:34&$2.38\pm0.2$&$2\pm1$&$2.93\pm0.3$&$10.07\pm0.3$&$-10.04\pm0.3$&$-0.80\pm0.08$&$-0.15\pm0.01$\\
Kr 1 &05:28:22&34:46:01&$1.98\pm0.2$&$3\pm2$&$2.69\pm0.3$&$09.89\pm0.3$&$-09.89\pm0.3$&$-0.32\pm0.03$&$-0.01\pm0.01$\\
Stock 8 &05:28:07&34:25:28&$1.98\pm0.20$&$2\pm1$&$2.69\pm0.3$&$09.89\pm0.3$&$-09.89\pm0.3$&$+0.31\pm0.03$&$-0.01\pm0.01$\\
\hline
\end{tabular}
\begin{list}{Table Notes.}
\item Cols. 2 and 3: Optimised central coordinates; Col. 4: $A_V$ in the cluster's central region. Col. 5: age, from 2MASS photometry. Col. 6: distance from the Sun. Col. 7: $R_{GC}$ calculated using $R_{\odot}=7.2$ kpc as the distance of the Sun to the Galactic centre. Cols. 8 - 10: Galactocentric components.
\end{list}
\label{tab4}
\end{center}
}
\end{table*}

\begin{figure}
\resizebox{\hsize}{!}{\includegraphics{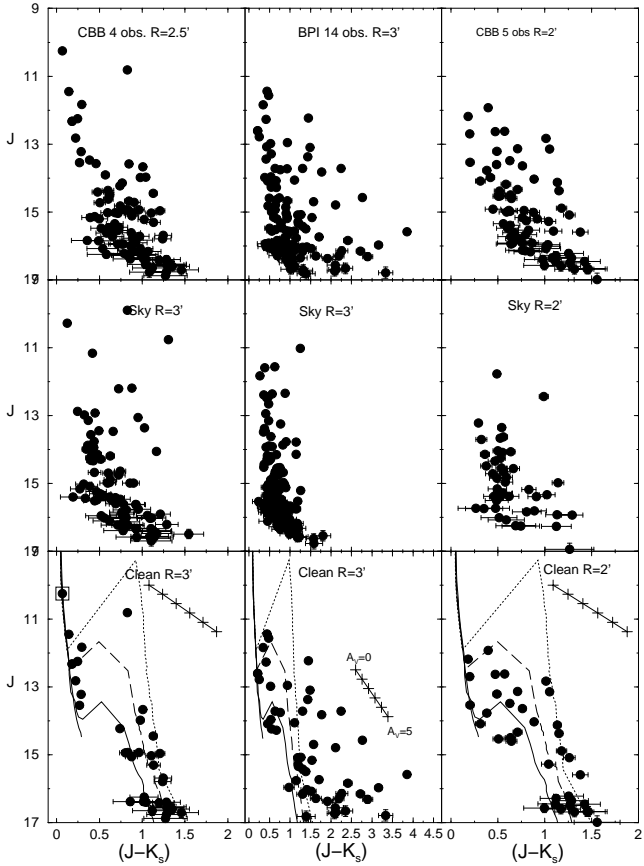}}
\caption[]{Same as Fig.~\ref{cmd1} for the clusters CBB 4, BPI 14, and CBB 5.}
\label{CBB7}
\end{figure}

\begin{figure}
\resizebox{\hsize}{!}{\includegraphics{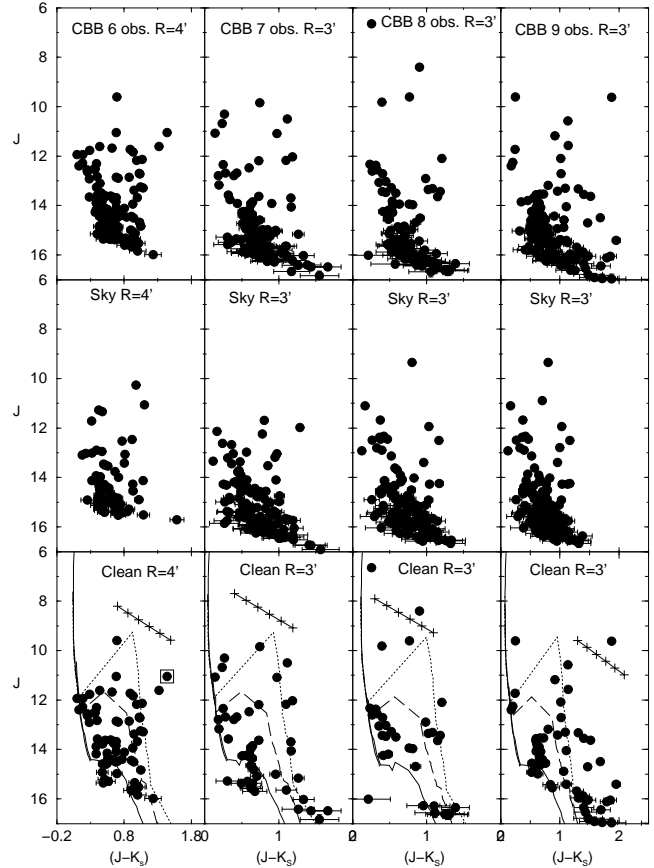}}
\caption[]{Same as Fig.~\ref{cmd1} for the clusters CBB 6, CBB 7, CBB 8, and CBB 9.}
\label{CBB9}
\end{figure}

\begin{table}
{\tiny
\caption{Young cluster indicators.}
\renewcommand{\tabcolsep}{1.0mm}
\renewcommand{\arraystretch}{1.4}
\begin{tabular}{lcccccccccc}
\hline
\hline
Cluster&IR stars&OB stars&$H_{\alpha}$&X-ray&DNe&BNe&RNe&Mcl/HII region\\
($1$)&($2$)&($3$)&($4$)&($5$)&($6$)&($7$) &($8$)&($9$)\\
\hline
BPI 14&x&&&&&&&x\\
CBB 3&x&x&&x&&&&\\
CBB 4&x&x&&&&&&x\\
CBB 5&x&&&&x&&&x\\
CBB 6&x&x&x&&&&&x\\
CBB 7&x&x&&&&&&x\\
CBB 8&x&x&&x&&&x&x\\
CBB 9&x&x&x&x&&&&x\\
FSR 722&x&x&&x&x&x&x&\\
FSR 734&x&x&x&x&&&&\\
FSR 777&x&x&x&x&&&&\\
FSR 780&x&x&x&x&&&&x\\
FSR 816&x&&x&&&x&&\\
FSR 817&&x&x&x&&&x&\\
FSR 833&x&x&x&&&&&\\
FSR 842&x&x&x&x&&&&\\
FSR 846&x&&x&&&&&\\
FSR 864&x&&x&&&&&\\
FSR 888&x&&&&x&&&x\\
FSR 890&x&x&x&&x&&&x\\
FSR 907&x&&x&&&&&x\\
FSR 925&x&&x&x&&&&\\
FSR 946&x&x&x&x&x&&&x\\
FSR 947&x&x&&&&&&\\
FSR 967&x&&x&&&&&\\
Kr 1&x&x&&x&&&&\\
Stock 8&x&x&x&x&&&&x\\
\hline
\end{tabular}
\begin{list}{Table Notes.}
\item Col.$(1)$ IR source, $(2)$ O or B stars, $(3)$ x-ray source, $(4)$ dark nebula, $(5)$ bright nebula and $(6)$ molecular cloud or HII region. From http://simbad.u-strasbg.fr/simbad/sim-fcoo and IPHAS.
\end{list}
\label{tab7}
}
\end{table}

\begin{figure}
\resizebox{\hsize}{!}{\includegraphics{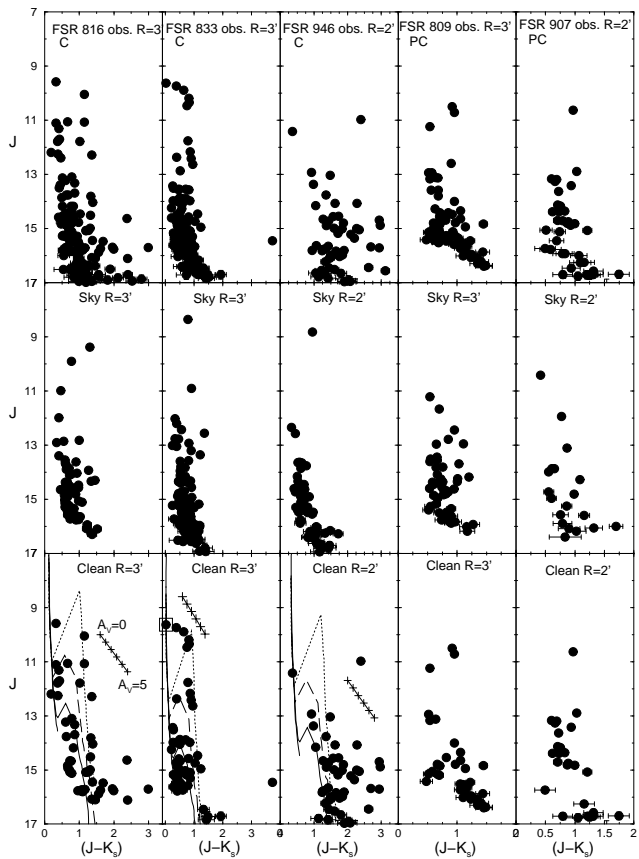}}
\caption[]{Same as Fig.~\ref{cmd1} for the clusters (FSR 816, FSR 833, and FSR 946) and the probable clusters (FSR 809 and FSR 907). We did not fit isochrones for the probable clusters. They require deeper observations.}
\label{cmd6}
\end{figure}

\begin{figure}
\resizebox{\hsize}{!}{\includegraphics{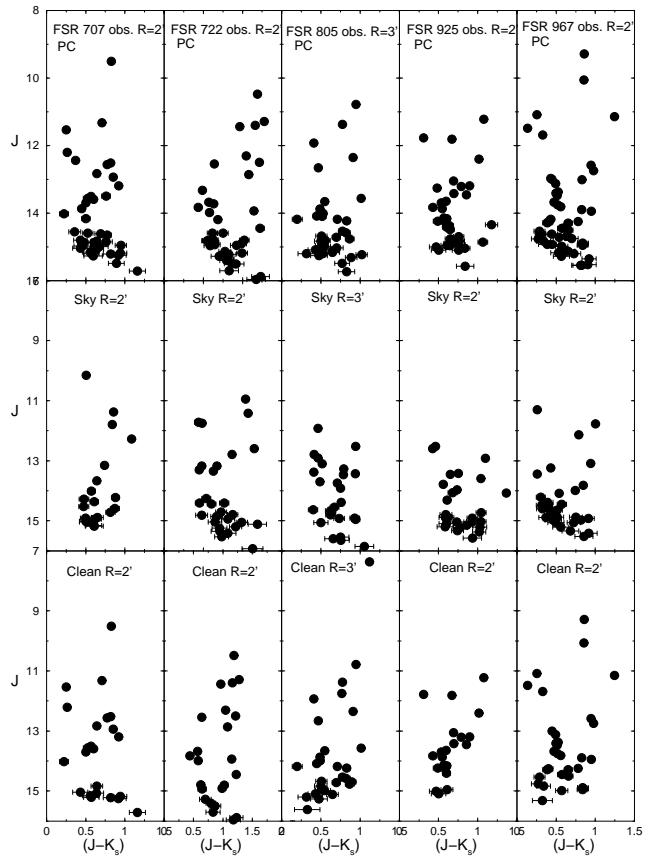}}
\caption[]{Same as Fig.~\ref{cmd1} for the probable clusters FSR 707, FSR 722, FSR 805, FSR 925, and FSR 967.}
\label{cmd7}
\end{figure}

\begin{figure}
\resizebox{\hsize}{!}{\includegraphics{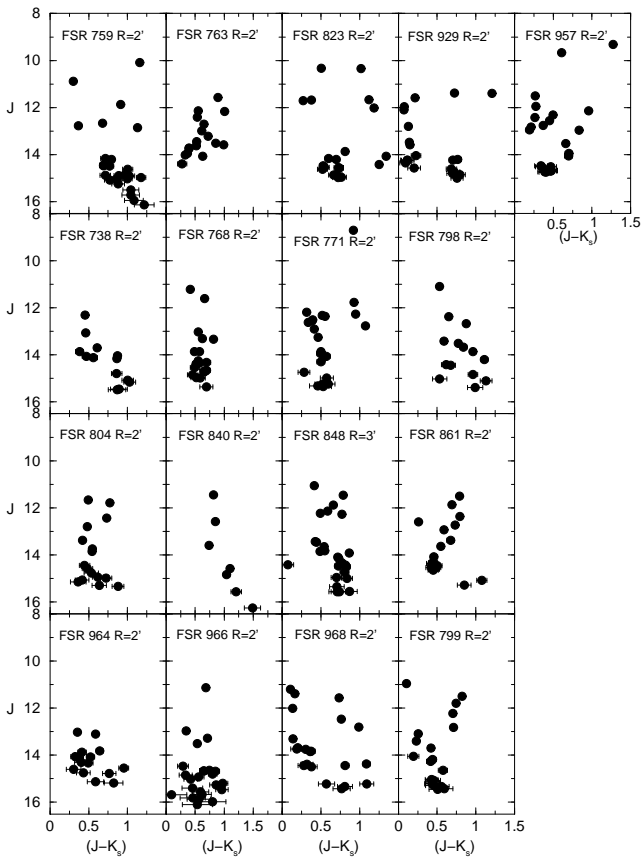}}
\caption[]{Decontaminated CMDs of the overdensities with lower probability of being star clusters.}
\label{lixo1}
\end{figure}

The CMDs for stars extracted from the central parts of the objects are shown in the upper panels of Figs.~\ref{cmd2} to \ref{lixo1}. The central part of a cluster is a region large enough to identify the main cluster evolutionary sequences in the CMD, but small enough to avoid significant contamination by field stars. This region is defined by inspections of the CMD and the RDP. The middle ones are nearly background extraction of equal area, and the bottom panels correspond to field decontaminated CMDs.
The fundamental parameters derived for the objects are showed in Table \ref{tab4}. The parameter errors have been estimated by displacing the best-fitting isochrone in colour and magnitude to the limiting point where the fit remains acceptable. We classified the overdensities FSR 707, FSR 722, FSR 805, FSR 809, FSR 907, FSR 925, and FSR 967 as probable clusters. These objects have, in general, less clear decontaminated CMD sequences than those confirmed as OCs. They also show irregular RDPs. The remaining 17 overdensities were classified by us as possible clusters, because their CMDs do not contain clear cluster sequences. We suggest deeper photometry to uncover the nature of the presently inferred probable and possible clusters. In Figs. \ref{cmd6} and \ref{cmd7} we show the decontaminated $J\times(J-K_s)$ CMDs for probable clusters, and in Fig.~\ref{lixo1}, the same for the possible ones.

In Table \ref{tab7} we add some young age indicators that helped us in determining the age of those clusters. 
The $H\alpha$ excesses were obtained from IPHAS \citep{Witham08, Drew05} and the emission-line star catalogue of \citet{Kohoutek99}. Additional indicators were obtained from SIMBAD\footnote{http://simbad.u-strasbg.fr/simbad}. $H\alpha$ and X-ray emission and IR excess may be correlated with the presence of circumstellar disks, accretion and stellar winds in PMS stars such as T Tauri and Herbig-Haro objects. These phenomena, together with the presence of OB stars, dark nebulae (DNe), reflection nebulae (RNe), Bright nebulae (BNe), H II regions, and molecular clouds, are consistent with the young age derived for the clusters.

For young clusters, the age determination is made through a combination of MS and PMS isochrones. PMS isochrones are especially important to estimate the age of clusters with poorly populated MSs.
Obviously, their distance determination depends on the assumed age, but the age that we adopted takes into account as well young age indicators (Table \ref{tab7}). They suggest that most clusters in our sample are embedded in the natal molecular cloud and some of them possibly have ongoing star formation. \citet{Lada03} inferred that  the duration of the embedded phase is 2-3 Myr and according to \citet{Hartmann01} stars older than $\sim5$ Myr are not found associated with molecular gas. The star formation timescale in spiral arms is $\sim1-4$ Myr and for a small cloud (40 pc), it is $\sim1$ Myr \citep{Elmegreen00, Ballesteros07, Tamburro08}. Some PMS stars in the present objects show IR-excess, but all PMS stars in a cluster lose their inner disks in $\sim6$ Myr and half of them can lose their disks in less than 3 Myr \citep{Haisch01}. The age spread in clusters with PMS stars is often assumed as $\sim10$ Myr \citep{Palla00}, but \citet{Jeffries11} suggest that the age dispersion in a young cluster is, in general, less than the median disk lifetime. The small merging or tidal destruction timescale for multiple clusters also suggest early age to ECs in groups in our sample. The relatively large distance uncertainty in the distance is a consequence of the age uncertainty.

\subsection{Colour-colour diagrams}
\label{sec:3.3}

\begin{figure}
\resizebox{\hsize}{!}{\includegraphics{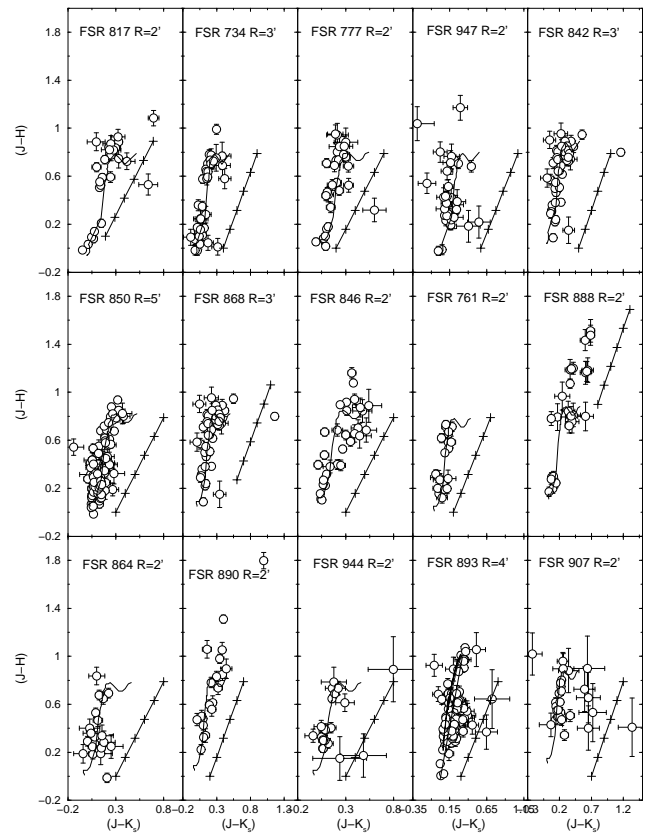}}
\caption[]{Colour-colour diagrams with the decontaminated photometry of the young confirmed clusters and a example of probable cluster (FSR 907). \citet{Siess00} isochrones and reddening vectors are used to characterise the PMS distribution.}
\label{color}
\end{figure}

\begin{figure}
\resizebox{\hsize}{!}{\includegraphics{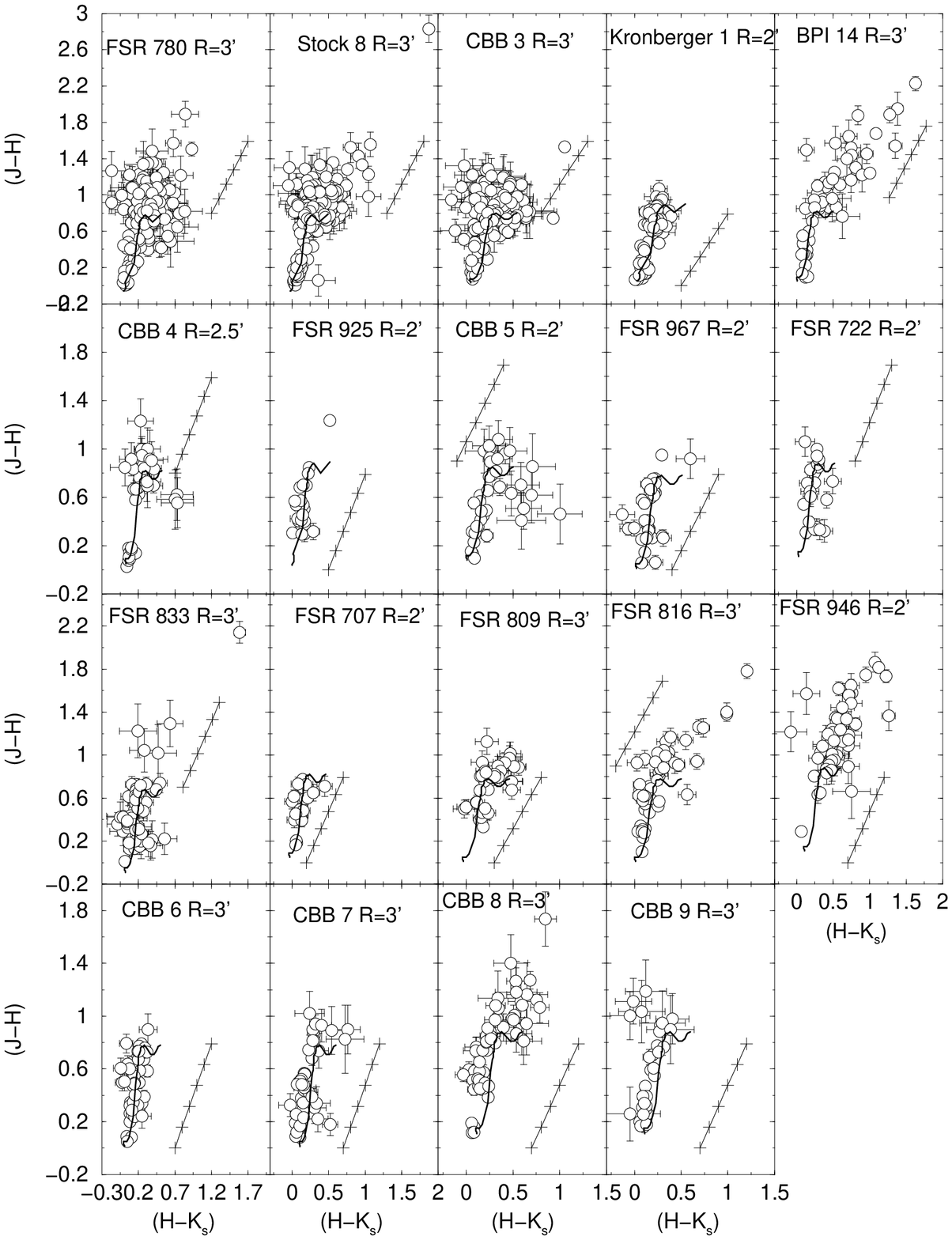}}
\caption[]{Same as Fig~\ref{color} for additional confirmed and probable clusters.}
\label{colour}
\end{figure}

Useful information on the nature and evolution of very young clusters, mainly about the emission of the stellar content in different regions of the spectrum, can be obtained with colour-colour diagrams. The colour-colour diagrams can be used to identify PMS stars and classify them. Since the present very young clusters include PMS stars, we show in Figs. \ref{color} and \ref{colour} the decontaminated near-IR colour-colour diagram $\jh\times\hk$ of the member stars, together with PMS tracks, set with the reddening values derived above, to estimate the age. Colour-colour diagrams of the present cluster sample show that a significant number of the stars appear to be very reddened, but the position of some stars, on the right side of the MS and PMS normal stars, suggest that they present a $K_s$-excess.  On the other hand, few appear to present an abnormal excess in $J$ and $H$. Generally, this excess is linked to stellar photospheric emission, but \citet{Cieza05} suggest a nonphotosperic nature for classical T Tauri stars and argue that J, H and $K_s$ excesses have a common source. The J-excess may be the cause of a negative value of $(J-K_s)$ for stars in some clusters' CMDs. 

We note the significant number of ECs characterised by a discontinuity between the distribution of the MS and PMS stars in the CMD (Figs. \ref{cmd2} - \ref{CBB7}). Until recently, such CMD features remained essentially inaccessible owing to the lack of field star decontamination \citep{Bonatto09a, Bonatto09b, Camargo09}.

\section{Cluster structure}
\label{sec:4}

The structure is analysed by means of the stellar RDP, defined as the projected number density around the cluster centre.
RDPs are built with stars selected after applying the colour magnitude (CM) filter to the observed photometry. This isolates the probable cluster sequences by excluding stars with deviant colours, thus enhancing the RDP contrast relative to the background \citep[e.g.][and references therein]{Bonatto07a}. However, field stars with colours similar to those of the cluster can remain inside the CM filter. They affect the intrinsic stellar RDP in a way that depends on the relative densities of field and cluster. The contribution of these residual field stars to the RDPs is statistically quantified by means of the comparison to the field. In practical terms, the use of the CM filters in cluster sequences enhances the contrast of the RDP with respect to the stellar field. The CM filters are shown in Figs.~\ref{cmd2} to \ref{cmd5} as the shaded area superimposed on the field-star decontaminated CMDs.

For simplicity we fit the RDPs with the function $\sigma(R)=\sigma_{bg}+\sigma_{0}/(1+(R/R_{core})^{2}$, where $\sigma_{bg}$ is the residual background surface density of stars, $\sigma_{0}$ is the central density of stars and $R_{core}$ is the core radius. The cluster radius ($R_{RDP}$) and uncertainty can be estimated by considering the fluctuations of the RDPs with respect to the residual background. $R_{RDP}$ corresponds to the distance from the cluster centre where RDP and comparison field become statistically indistinguishable. This function, applied to star counts, is similar to that introduced by \citet{King62} to describe the surface-brightness profiles in the central parts of globular clusters. To minimise degrees of freedom $\sigma_{bg}$ is measured in the comparison field and kept fixed.

\begin{figure}
\resizebox{\hsize}{!}{\includegraphics{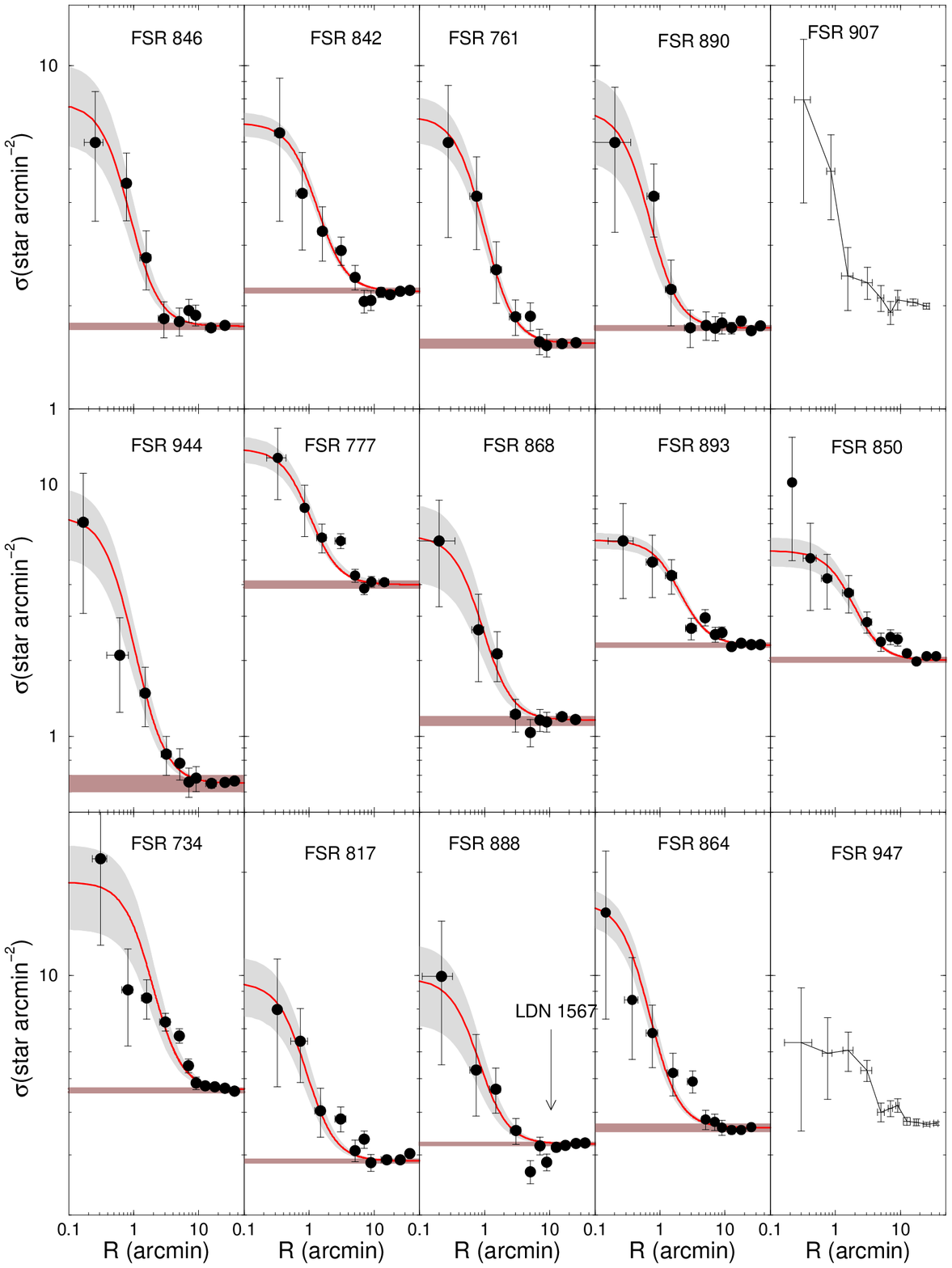}}
\caption[]{Radial density profiles for the confirmed clusters and the probable cluster FSR 907.}
\label{rdp1}
\end{figure}

\begin{figure}
\resizebox{\hsize}{!}{\includegraphics{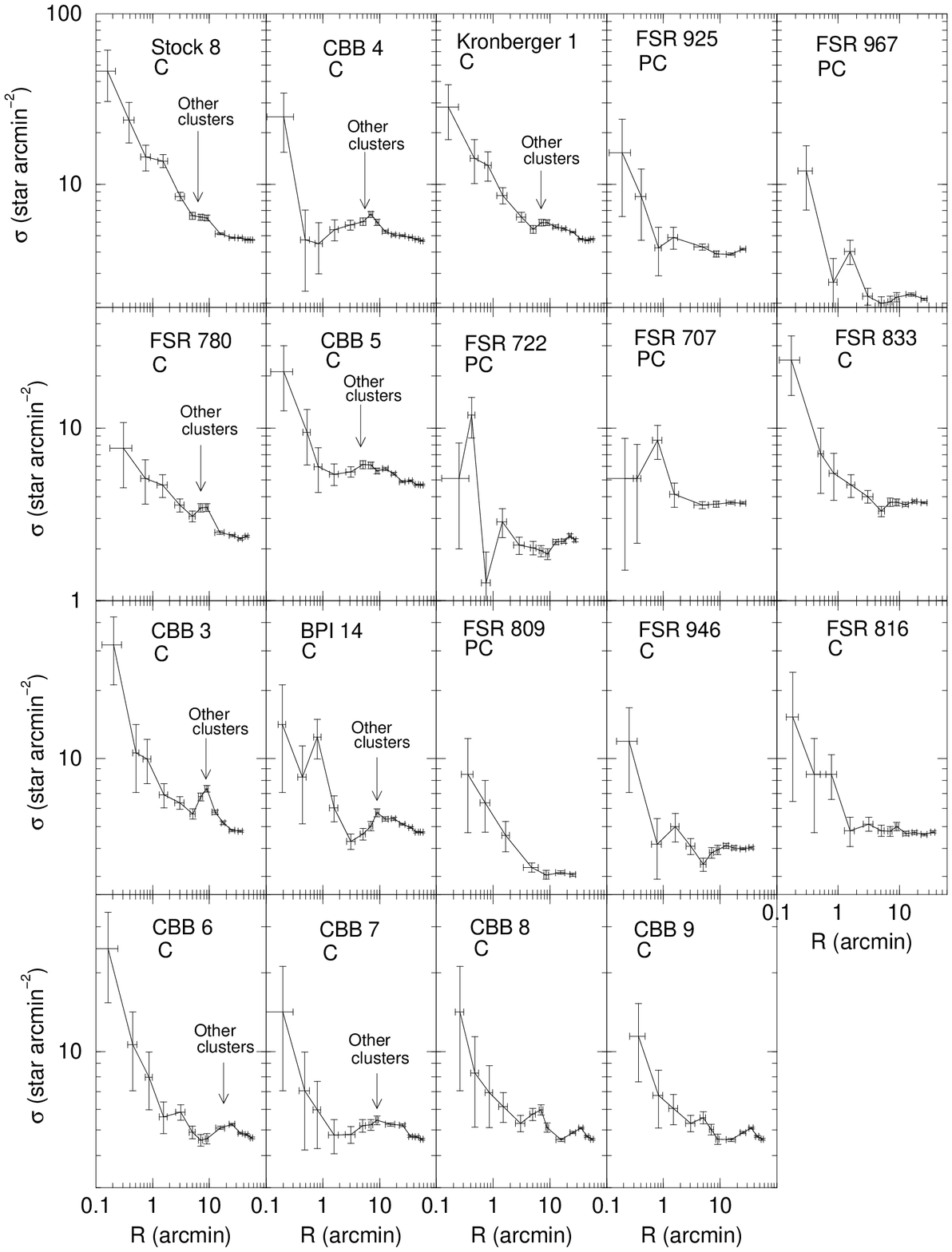}}
\caption[]{Radial density profiles for the confirmed (C) and probable clusters (PC).}
\label{rdp2}
\end{figure}

\begin{figure}
\resizebox{\hsize}{!}{\includegraphics{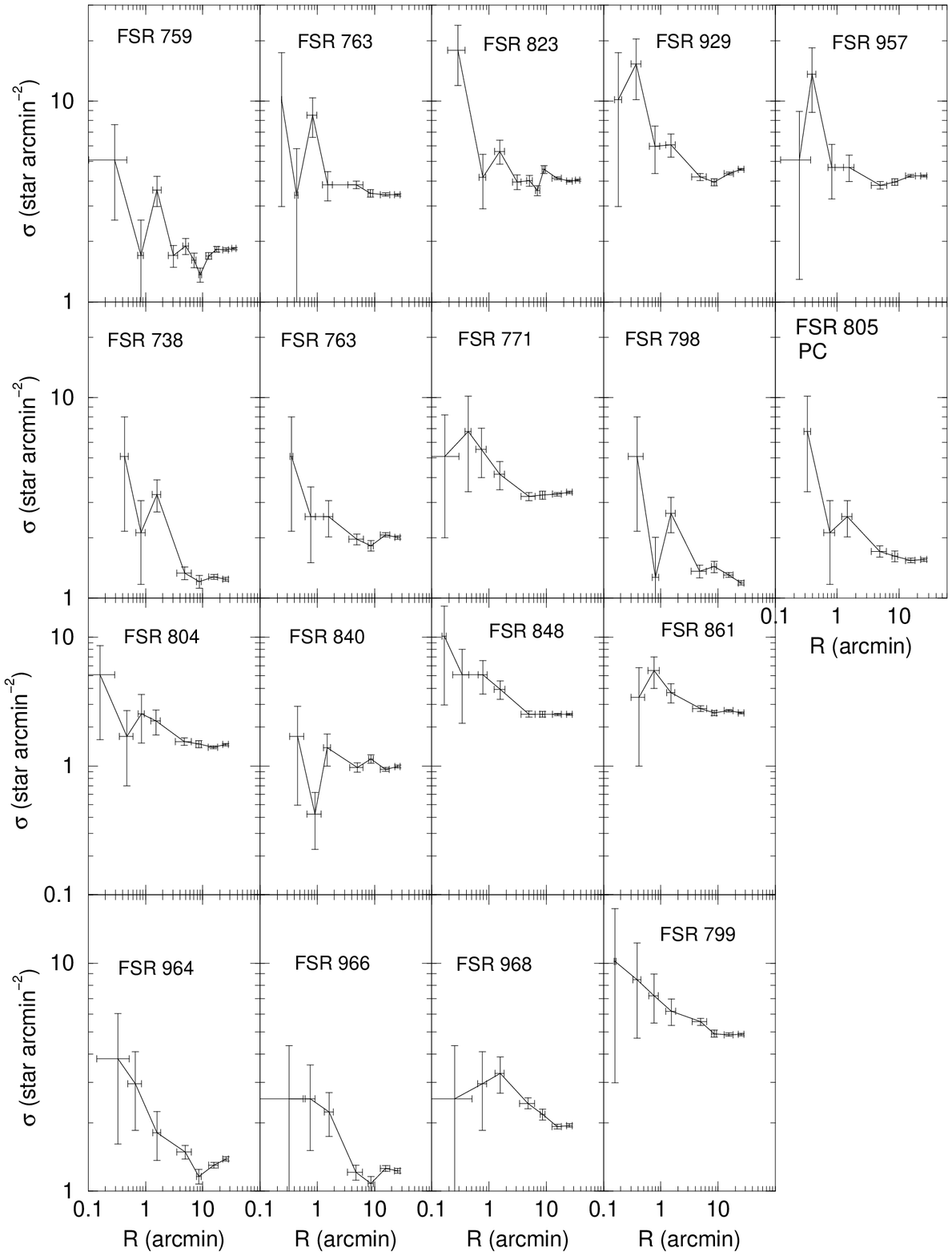}}
\caption[]{Radial density profiles for the overdensities that have lower probability of being OCs and the probable cluster FSR 805.}
\label{lixo2}
\end{figure}

Structural parameters derived are shown in Table~\ref{tab5} and the RDPs are in Figs.~\ref{rdp1}, \ref{rdp2}, and \ref{lixo2}.

\begin{table*}
{\footnotesize
\begin{center}
\caption{Structural parameters for 13 confirmed clusters.}
\renewcommand{\tabcolsep}{0.9mm}
\renewcommand{\arraystretch}{1.3}
\begin{tabular}{lrrrrrrrrrrr}
\hline
\hline
Cluster&$(1')$&$\sigma_{0K}$&$\sigma_{bg}$&$R_{core}$&$R_{RDP}$&$\sigma_{0K}$&$\sigma_{bg}$&$R_{core}$&$R_{RDP}$&${\Delta}R$&CC\\
&($pc$)&($*\,pc^{-2}$)&($*\,pc^{-2}$)&($pc$)&($pc$)&($*\,\arcmin^{-2}$)&($*\,\arcmin^{-2}$)&($\arcmin$)&($\arcmin$)&($\arcmin$)&\\
($1$)&($2$)&($3$)&($4$)&($5$)&($6$)&($7$)&($8$)&($9$)&($10$)&($11$)&($12$)\\
\hline
FSR 734 &$0.76$&$24.6\pm5.4$&$8.0\pm0.14$&$1.05\pm0.21$&$7.22\pm1.4$&$14.2\pm3.13$&$4.65\pm0.08$&$1.38\pm0.28$&$9.5\pm1.5$&$20-60$&$0.93$\\
FSR 761 &$0.74$&$10.17\pm1.82$&$2.83\pm0.09$&$0.50\pm0.07$&$3.7\pm0.7$&$5.57\pm1.0$&$1.55\pm0.05$&$0.68\pm0.10$&$5.3\pm1.0$&$20-30$&$0.98$\\
FSR 777 &$0.78$&$16.2\pm2.1$&$5.2\pm0.11$&$0.59\pm0.06$&$4.3\pm0.8$&$9.85\pm1.3$&$4.1\pm0.07$&$0.76\pm0.08$&$5.5\pm1.0$&$8-18$&$0.98$\\
FSR 817 &$0.68$&$14.4\pm4.1$&$6.3\pm0.13$&$0.45\pm0.07$&$4.08\pm1.02$&$6.68\pm1.9$&$2.9\pm0.06$&$0.67\pm0.1$&$6.0\pm1.5$&$20-30$&$0.96$\\
FSR 842 &$0.62$&$11.96\pm0.52$&$5.72\pm0.10$&$0.65\pm0.06$&$4.34\pm0.06$&$4.60\pm0.2$&$2.2\pm0.04$&$1.05\pm0.09$&$7.0\pm2.0$&$20-40$&$0.99$\\
FSR 846 &$0.72$&$11.63\pm4.05$&$3.35\pm0.13$&$0.43\pm0.10$&$3.24\pm$&$6.03\pm2.1$&$1.74\pm0.05$&$0.60\pm0.15$&$4.5\pm1.5$&$20-30$&$0.92$\\
FSR 850 &$0.79$&$5.5\pm1.1$&$2.5\pm0.06$&$1.2\pm0.03$&$7.9\pm2.4$&$3.45\pm0.7$&$2.02\pm0.04$&$1.51\pm0.26$&$10.0\pm3.0$&$20-30$&$0.96$\\
FSR 864 &$0.84$&$15.05\pm0.95$&$5.10\pm0.07$&$0.40\pm0.06$&$5.04\pm0.84$&$12.64\pm0.67$&$3.60\pm0.05$&$0.48\pm0.07$&$6.0\pm1.0$&$20-30$&$0.99$\\
FSR 868 &$0.79$&$8.2\pm3.4$&$1.8\pm0.10$&$0.48\pm0.15$&$3.95\pm1.58$&$5.11\pm2.1$&$1.16\pm0.06$&$0.61\pm0.19$&$5.0\pm2.0$&$20-40$&$0.90$\\
FSR 888 &$0.80$&$10.3\pm3.9$&$5.0\pm0.06$&$0.49\pm0.2$&$3.2\pm0.80$&$6.57\pm2.5$&$3.23\pm0.04$&$0.62\pm0.2$&$4.0\pm1.0$&$20-40$&$0.90$\\
FSR 890 &$0.75$&$10.1\pm3.9$&$3.1\pm0.05$&$0.36\pm0.1$&$2.25\pm0.75$&$5.69\pm2.2$&$1.72\pm0.03$&$0.49\pm0.2$&$3.0\pm1.5$&$20-40$&$0.85$\\
FSR 893 &$0.32$&$36.13\pm3.90$&$22.46\pm0.49$&$0.51\pm0.08$&$2.88\pm0.64$&$3.70\pm0.4$&$2.30\pm0.05$&$1.61\pm0.24$&$9.0\pm2.0$&$20-40$&$0.96$\\
FSR 944 &$0.70$&$13.87\pm2.65$&$1.33\pm0.20$&$0.39\pm0.07$&$3.85\pm1.1$&$6.80\pm1.3$&$0.65\pm0.1$&$0.56\pm0.1$&$5.5\pm1.5$&$20-40$&$0.97$\\
\hline
\end{tabular}
\begin{list}{Table Notes.}
\item Col. 2: arcmin to parsec scale. To minimise degrees of freedom in RDP fits with the King-like profile (Sect.~\ref{sec:4}), $\sigma_{bg}$ was kept fixed (measured in the respective comparison fields) while $\sigma_{0}$ and $R_{core}$ were allowed to vary. Col. 11: comparison field ring. Col. 12: correlation coefficient.
\end{list}
\label{tab5}
\end{center}
}
\end{table*}

\begin{table*}
\caption[]{Stellar mass estimate for star clusters with PMS.}
\label{tab8}
\renewcommand{\tabcolsep}{2.6mm}
\renewcommand{\arraystretch}{1.25}
\begin{tabular}{cccccccccccc}
\hline\hline
&&\multicolumn{4}{c}{MS}&&\multicolumn{2}{c}{PMS}&&\multicolumn{2}{c}{$MS+PMS$}\\
\cline{3-6}\cline{8-9}\cline{11-12}
Cluster&&$\Delta\,m_{MS}$&&$N$&$M$   &&$N$&$M$ &&$N$&$M$\\
     && ($M_\odot$)    &      &(stars) &($M_\odot$)&&(stars)  &($M_\odot$)&&(stars) &($M_\odot$)     \\
(1)&&(2)&&(3)&(4)&&(5)&(6)&&(7)&(8)\\
\hline
FSR 734&&0.18-95.0&&$751\pm515$&$905\pm319$&&$133\pm23$&$80\pm14$&&$884\pm538$&$985\pm333$\\
FSR 761&&1.30-11.0&&$148\pm75$&$460\pm266$&&$49\pm14$&$29\pm8$&&$197\pm89$&$489\pm274$\\
FSR 777&&2.90-17.0&&$37\pm22$&$264\pm161$&&$87\pm15$&$52\pm9$&&$124\pm37$&$316\pm170$\\
FSR 817&&2.50-11.0&&$113\pm15$&$197\pm46$&&$94\pm14$&$56\pm8$&&$207\pm29$&$253\pm54$\\
FSR 842&&1.50-17.0&&$219\pm125$&$280\pm64$&&$79\pm11$&$47\pm11$&&$298\pm136$&$327\pm75$\\
FSR 846&&2.30-19.0&&$281\pm176$&$294\pm95$&&$47\pm10$&$28\pm6$&&$328\pm186$&$341\pm101$\\
FSR 850&&0.95-11.0&&$2790\pm2060$&$1220\pm396$&&$51\pm11$&$31\pm7$&&$2841\pm2456$&$1251\pm403$\\
FSR 864&&1.10-13.0&&$796\pm558$&$413\pm110$&&$46\pm8$&$28\pm16$&&$842\pm566$&$441\pm126$\\
FSR 868&&1.70-6.25&&$116\pm2$&$174\pm5$&&$42\pm9$&$25\pm5$&&$158\pm11$&$199\pm115$\\
FSR 888&&2.70-11.0&&$249\pm175$&$145\pm41$&&$45\pm7$&$27\pm4$&&$294\pm182$&$172\pm45$\\
FSR 890&&2.30-11.0&&$288\pm194$&$199\pm49$&&$33\pm6$&$20\pm4$&&$321\pm200$&$219\pm53$\\
FSR 944&&1.90-9.75&&$298\pm187$&$251\pm61$&&$20\pm5$&$12\pm3$&&$318\pm192$&$263\pm65$\\
\hline
\end{tabular}
\begin{list}{Table Notes.}
\item Col. 2: MS mass range. Cols. 3-6: stellar content of the MS and PMS stars. Cols. 7-8: total (MS+PMS) stellar content.
\end{list}
\end{table*}

\begin{table*}
\caption[]{Stellar mass estimate for the old open cluster FSR 893}
\label{mass2}
\renewcommand{\tabcolsep}{3.8mm}
\renewcommand{\arraystretch}{1.25}
\begin{tabular}{ccccccccccc}
\hline\hline
&&\multicolumn{5}{c}{Observed in the CMD}&&\multicolumn{2}{c}{Extrapolated}\\
\cline{3-7}\cline{9-10}
Cluster&&$\Delta\,m_{MS}$&$N_{MS}$&$M_{MS}$   &$N_{evol}$&$M_{evol}$ &&$N$&$M$\\
     && ($M_\odot$)         &(stars) &($M_\odot$)&(stars)   &($M_\odot$)&&(stars) &($M_\odot$)     \\
(1)&&(2)&(3)&(4)&(5)&(6)&&(7)&(8)\\
\hline
FSR 893&&0.35-1.50&$646\pm158$&$357\pm73$&$11\pm7$&$15\pm10$&&$2510\pm2420$&$677\pm473$\\

\hline
\end{tabular}
\begin{list}{Table Notes.}
\item Col. 2: MS mass range. Cols. 3-6: stellar content of the MS and evolved stars. Cols. 7-8: stellar content extrapolated to $0.08\,M_{\odot}$.
\end{list}
\end{table*}

\section{Mass estimates}
\label{Mass}

Due to the relatively large distance of the OCs and 2MASS photometry limitations, we do not have access to the whole stellar mass range of these clusters. Therefore, for the old OC FSR 893 we use Kroupa's (2001) mass function to estimate the total stellar mass, down to the H-burning mass limit ($0.08\,M_{\odot}$). The estimated mass is shown in Table~\ref{mass2}.

For MS stars in young clusters we simply count stars in the CMDs (within the region $R<R_{RDP}$), and sum their masses as estimated from the mass-luminosity relation implied by the respective isochrone solution (Sect. \ref{sec:3.2}). Subsequently, we count the number of PMS stars and multiply by an average mass value adopted for these stars to estimate the mass  within the PMS. Assuming that the mass distribution of the PMS stars also follows Kroupa's (2001) MF, the 
average PMS mass - for masses within the range $0.08\la m(\ms)\la7$ is $<m_{PMS}>\approx0.6\ms$. The estimated mass (Table~\ref{tab8}) should be taken as a lower limit. 

\section{Discussion}
\label{sec:7}
\subsection{General}
\label{sec:7.1}

The early EC's structure reflects the underlying fractal structure in the dense molecular gas in which they are formed \citep{Beech87, Lada03}, but the dynamical evolution may erase this shape leading these objects to a smooth structure. Previous works suggest that ECs are not virialised and, as a consequence, are sub-structured with a RDP presenting bumps and dips comparable to the field fluctuation. Because of this, they cannot be fitted by King's profile that describes the structure of clusters close to spherical symmetry and centrally concentrated \citep{Soares05, Gutermuth05, Camargo11}. However, after the gas expulsion, surviving OCs are probably not virialised, so the bumps and dips may be the result of a non relaxed cluster before the gas expulsion, or a consequence of the gas expulsion. An additional explanation is the presence of other clusters in the neighbourhood.

Objects like Stock 8, FSR 780, CBB 3, and neighbours (Fig.~\ref{rdp2}) are possibly examples of out-of-equilibrium star clusters. In addition, the presence of other objects creates bumps in the RDP (Fig.~\ref{rdp2}). Therefore, their structural parameters cannot be derived by a King law, but it does not mean that the RDP does not provide information on the cluster structure. For example, it can be useful to differentiate physical systems from field fluctuations.  Furthermore, the CMD morphology may provide clues to their nature \citep{Bica11}. However, deeper observations  of  the less-populated clusters with irregular RDPs could solve this problem by checking if the irregularities are intrinsic to the cluster structure or result from the photometric constraints. 

Most objects confirmed as clusters in this work are ECs.
FSR 893 is the unique old OC in our sample, with age of $\sim3$ Gyr.
The position of the bluest stars in the CMD of FSR 893 (Fig.~\ref{FSR893}) are consistent with blue straggler.  

Some ECs present stars with IR-excess such as FSR 842 and BPI 14, but other objects possibly present J-excess resulting in a negative value of ($J-K_s$), e.g., FSR 817 and FSR 734. FSR 842 is a young cluster close to the local Arm and projected 18' from the centre of the small molecular cloud $\sharp\,64$ of \citet{Kawamura98}. It presents a very reddened emission-line B star in the central region. As shown in Figs. \ref{cmd2} and \ref{color} PMS stars do not present a significant infrared excess, one exception is the B star that is heavily affected by IR-excess. It also presents $H_{\alpha}$-excess, and is located in the colour-colour diagram (Fig. \ref{color}) in the loci expected for classical Be stars \citep{Hernandez05}. For clusters with age $\sim5$ Myr the disk frequency in intermediate mass stars is often higher than for low mass stars \citep{Kennedy09, Hernandez11}. However, the timescale of the disk dissipation for intermediate mass stars is $\sim3$ Myr. Probably, stellar collisions produce the dust triggering a second generation of planetary disks.  

Some objects classified as probable clusters exhibit evidence of being young clusters (Table \ref{tab7}). 

\begin{figure}
\resizebox{\hsize}{!}{\includegraphics{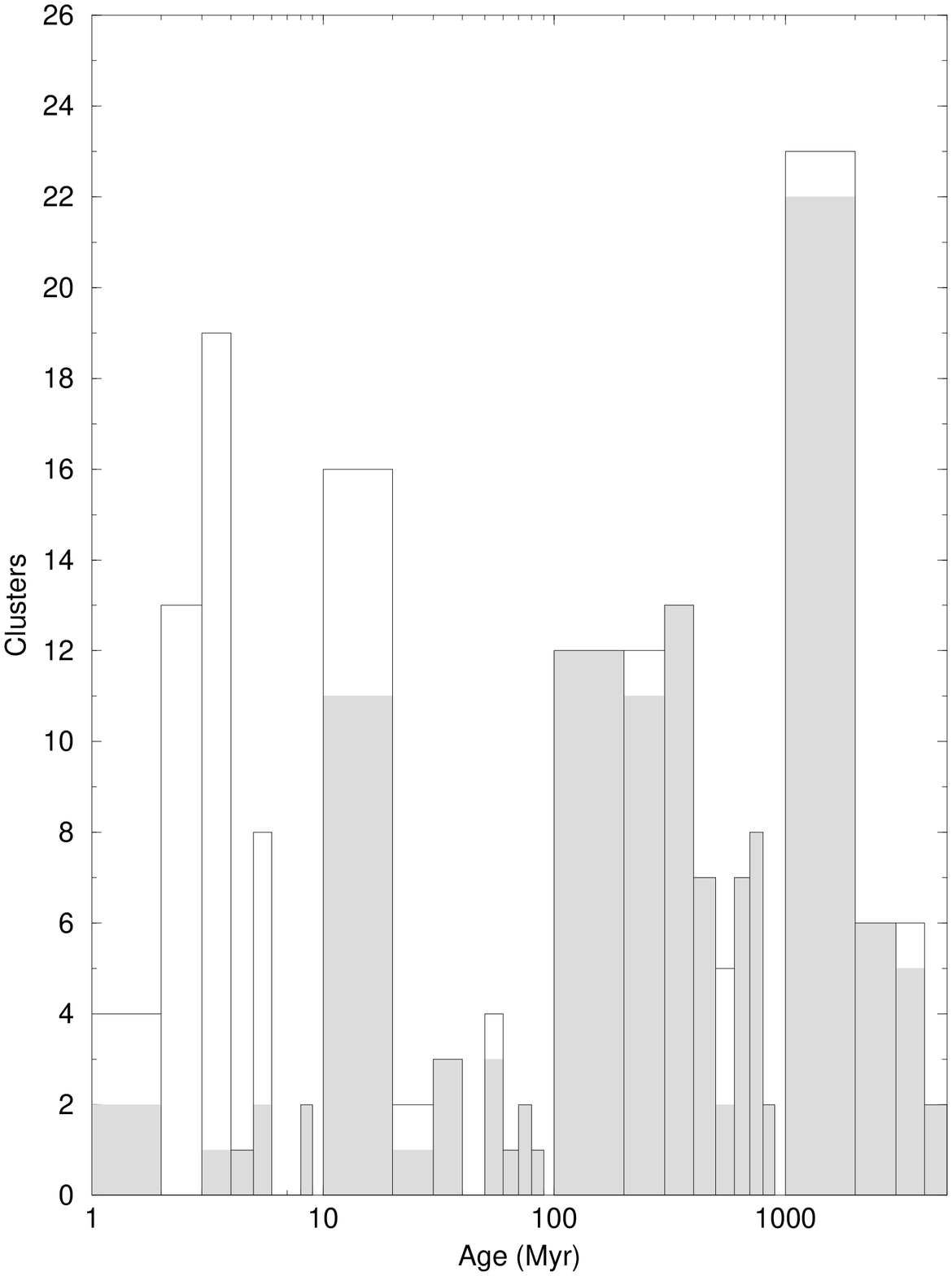}}
\caption[]{The gray-shaded histogram presents the age distribution for clusters in the latest version of the DAML02 catalogue, except for clusters in \citet{Camargo10}, and the continuous black line histogram represents DAML02 coupled with our contribution \citep[][and the present work]{Camargo10, Camargo11} to the anticentre clusters.}
\label{Histograma}
\end{figure}

The latest version of DAML02 (January, 2012) presents 206 clusters towards the Galactic anticentre, 137 of them with age. Fig.~\ref{Histograma} shows histograms with the age distribution of clusters in DAML02 after removing the contribution of \citet{Camargo10}, and DAML02 coupled with our contribution \citep[][and the present work]{Camargo10, Camargo11} to the anticentre clusters. Based on this distribution we deduce that $\sim80\%$  of the clusters in this region are dissolved in less than 1 Gyr, and estimate an average age of $\sim570$ Myr for the clusters in the anticentre.
In this sense, our results increase the number of clusters with derived parameters towards the anticentre significantly ($\sim38\%$), especially young ones. However, the number of clusters younger than 10 Myr represent less than $26\%$ of all clusters towards the anticentre. 

\subsection{Clusters discovered in the present work}

We discovered 6 new clusters (CBB 3, CBB4, CBB5, CBB 6, CBB 7 and CBB 9) in the Stock 8 neighbourhood (Fig. \ref{IC}) and one (CBB 8) in the nebula Sh2-229. Figs. \ref{CBB5}, \ref{CBB7}, and \ref{CBB9} show the CMDs and Fig.~\ref{rdp2} the RDPs of these objects, the derived parameters are shown in Tables \ref{tab4} and \ref{tab5}.

They are newborn, low mass, and poorly populated clusters with a well-defined core and RDPs that do not follow a King's law (Fig.~\ref{rdp2}). CBB 3 ($R\sim5'$), CBB 4 ($R\sim3'$), and CBB 5($R\sim3'$) are close to Stock 8. CBB 6 with a radius of $\sim5'$ presents an $H_{\alpha}$ emitter B star classified as YSO (Fig.~\ref{IC}) in SIMBAD. CBB 7 is a small cluster ($R\sim3'$) close to FSR 777. CBB 8 is partly embedded in the nebula Sh2-229 and presents an O star in the neighbouring field. The CBB 8 structure (Fig.~\ref{rdp2}) and environmental conditions suggest that this object is in the process of evaporation \citep{Lada03, Bastian06}. CBB 9 with a radius of $\sim6'$ includes an O star. 

\subsection{Sequential star formation in Aur OB2?}

FSR 780 is a young cluster located near Stock 8 in IC 417 (Sh2-234) within Aur OB2.  According to \citet{Mel'nik95} Aur OB2 is located at 2.68 kpc from the Sun. \citet{Fich84} estimated for IC 417 a kinematic distance of $2.3\pm0.7$ kpc. On the other hand, \citet{Jose08} derived an age younger than 2 Myr for Stock 8 and a distance from the Sun of $2.05\pm0.10$ kpc with a radius of $\approx6'$. They indicated an enhancement in stellar density at $\approx7'.5$ from the centre.   

The presently discovered clusters appear to be linked to Stock 8, FSR 777, Kronberger 1, and BPI 14 in IC 417.
We suggest that BPI 14 \citep{Borissova03} is associated with IC 417 supporting \citet{Jose08} results. The presence of other clusters is indicated as bumps in the RDP of each cluster candidate, which explains the enhancement observed by \citet{Jose08} in the field of Stock 8. The RDPs of these objects show independent peaks supporting the interpretation that they are distinct compact clusters. Examples of such structures are discussed by \citet{Feigelson11} for clusters in the Carina complex and by \citet{Camargo11} for ECs in Sh2-235.

\begin{figure}
\resizebox{\hsize}{!}{\includegraphics{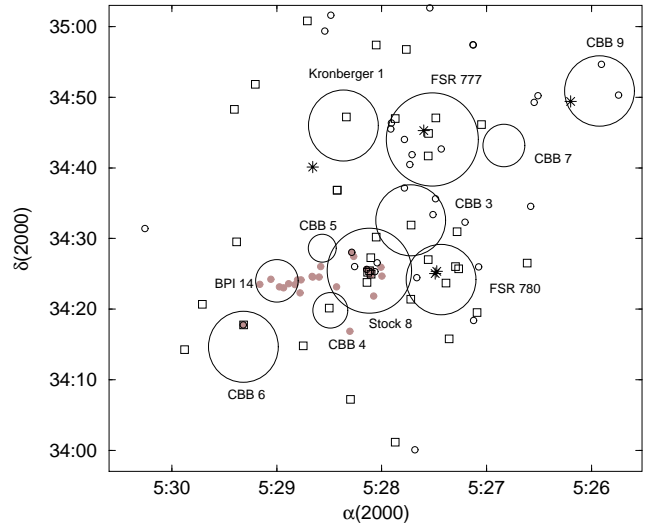}}
\caption[]{Angular distribution of clusters (large circles) and massive stars concentrating in Aur OB2. The filled circles are O stars, the squares are B stars, open circles are $H_{\alpha}$ emitters and brown circles are YSOs.}
\label{IC}
\end{figure}

The angular distribution of clusters, massive stars, nebulae, and young stellar objects (YSOs) support a scenario of sequential star formation triggered by massive stars in FSR 780 (Fig.~\ref{IC}). However, the age gradient is comparable to the uncertainty in ages.

The stellar density distribution in Aur OB2 shows small peaks coinciding mainly with massive stars (Fig. 19). The formation of massive stars in early cluster phases may destroy them, because winds blow out gas and dust. However, as most massive stars form in clusters \citep{Lada03, deWit05, Fall05, Weidner06} the occurrence of isolated ones may be related to the infant mortality rate\footnote{If these stars are actually linked with this star forming region.}. It is possible that massive star formation within an OB association occurs in \textit{clumps} that merge forming massive clusters, or disrupt them becoming field OB stars.

Recent works suggest that stars may be formed in low density stellar groups (LDSGs) that disperse in the field
without requiring gas expulsion \citep{Bastian11, Moeckel11, Kruijssen12}.
LDSGs are generally distributed along filamentary structures of gas and dust \citep{Gutermuth08, Myers09}. Filamentary structures are sites of turbulent gas motion, which seems to
favour the formation of individual stars or LDSGs,
with turbulence enhancing gas densities, and decreasing the Jeans mass
\citep{Low02, Vazquez03, Clark05}. \citet{Bressert10} estimate that only $\sim26\%$ of the YSOs near the Sun
 are located in ECs.
This scenario explains  probable young stars outside clusters, and the continuous distribution of YSOs between Stock
8 and BPI 14, supporting sequential star formation (Fig.~\ref{IC}).

The stellar distribution in Aur OB2 suggests that  formation in clustered environments \citep[][and similar works]{Kroupa01, Lada03} and in LDSGs
 may be part of a global scenario, where star formation occurs in groups with a continuous range of stellar densities depending on the natal gas density, with dense \textit{clumps} forming bound clusters and low density gas producing  LDSGs \citep[][and reference therein]{Elmegreen08}. ECs with the gas centrally concentrated are more dependent on the SFE than sub-structured ones, which are less affected by gas expulsion. LDSGs probably result unbound, or merge forming a cluster. The latter is a way to form clusters in low density molecular gas and contributes to explain the low SFE in GMCs. These groups may contribute to forming associations of
clusters, since filamentary structures are often found near ECs. Nevertheless, these stars might be ejected from
the clusters by dynamical effects \citep[][and reference therein]{Gvaramadze11}.

Sequential star formation is also possible for FSR 777, Kronberger 1, and CBB 7. In this scenario the star formation might have been triggered by massive stars in FSR 777 and the O star below Kronberger 1 (Fig.~\ref{IC}). \citet{Kronberger06} estimated for Kronberger 1  a distance from the Sun of 1.9 kpc and age of $\sim32$ Myr, but we argue, based on the very young age indicators (Table~\ref{tab7}) and CMD (Fig.~\ref{cmd1}) that this object is newborn and located in the Perseus arm.

Recently, we investigated a group of compact ECs related to four H II regions (Sh2-235, Sh2-233, Sh2-232, and Sh2-231) that is possibly developing sequential star formation \citep{Camargo11}. This region presents some cluster pairs like in IC 417. 

\begin{figure*}
\resizebox{\hsize}{!}{\includegraphics{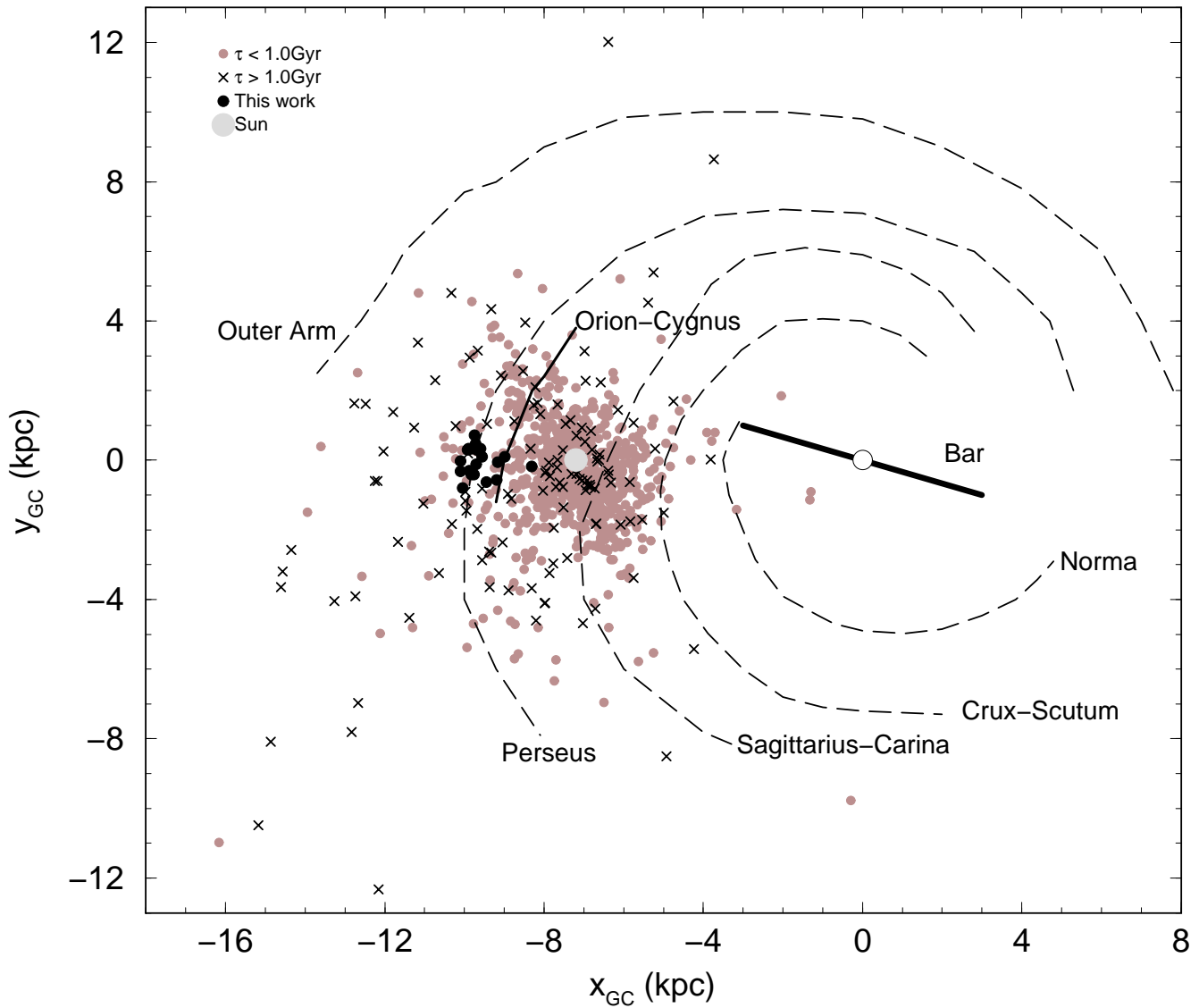}}
\caption[]{Spatial distribution of the confirmed clusters (filled circles) compared to the WEBDA OCs with ages younger than 1 Gyr (brown circles) and older than 1 Gyr (crosses). The schematic projection of the Galaxy is seen from the north pole \citep{Momany06}, with 7.2 kpc as the Sun's distance to the Galactic centre.}
\label{spiral}
\end{figure*}

\subsection{OB associations}

OB associations are often sub-structured consisting of several sub-groups. An irregular GMC may form massive stars simultaneously and their winds and/or supernova explosions may produce a second generation of massive stars propagating the star formation and forming star clusters with a small age spread \citep{Elmegreen77}. The timescale required for a complete star formation process in a star complex is 10 to 20 Myr \citep{Fuente08}. 

The clusters Stock 8, FSR 780, CBB 3, CBB 4, CBB 5, BPI 14, Kronberger 1, and FSR 777 form an association of clusters related to the HII region IC 417 in the Aur OB2 association. Such cluster structure has similarities with the clusters and groups in the Carina Complex \citet{Feigelson11}. According to the results for these clusters (Table \ref{tab4}), Aur OB2 is located in the Perseus arm. Also, \citet{Bonatto09b} derived a distance of $2.4\pm0.3$ kpc and age of about 10 Myr for NGC 1931 that is embedded in Sh2-237 within Aur OB2. \citet{Moffat79} found $d_{\odot}=1.8$ kpc, \citet{Pandey86} and \citet{Bhatt94}  2.2 kpc, and \citet{Chen04} found 3.1 kpc. A more populous EC in this region is NGC 1893 for which \citet{Tapia91} estimated an age of about 4 Myr and distance of 4.3 kpc. \citet{Sharma07} locate NGC 1893 at $3.25\pm0.02$ kpc and find that it is younger than 3 Myr. The distance derived for the cluster CBB 8 suggests that the nebula Sh2-229 belongs to Perseus arm, despite the uncertainty in the distance determination.

\citet{Camargo11} derived parameters for ECs related to the H II regions Sh2-231, Sh2-232, Sh2-233, and Sh2-235 in the direction of Aur OB1 and near the Aur OB2 borders. Within uncertainties these objects may belong to the Perseus arm. \citet{Straizys10} suggest that Sh2-231 may belong to the Perseus arm, but they found a distance of 1.3 kpc for the other nebulae, which agree with the distance estimated by \citet{Humphreys78} for Aur OB1. This value is often assumed as the distance of objects in Aur OB1.  

FSR 888 and FSR 890 are embedded in the nebula Sh2-249 (LBN 188.69+04.25) in Gem OB1. Our parameters for these objects suggest that Sh2-249 and consequently Gem OB1 are close to the Perseus arm.
The distance to Gem OB1 has been estimated to be 1.2 - 2 kpc \citep{Haug70, Humphreys78}. \citet{Carpenter95} found a distance to Gem OB1 of 1.5 kpc and 2 kpc to a GMC towards this association. 

\citet{Russeil07} estimated a distance of $2.46\pm0.16$ kpc to the complex 192.5-0.1 that is composed of Sh2-254, Sh2-255, Sh2-256, Sh2-257 and Sh2-258 \citep{Chavarria08, Ojha11}. In addition, \citet{Reid09} argue that Sh2-252 is located in the Perseus arm. \citet{Bonatto11} derived a distance of about 1.5 kpc for clusters in Sh2-252.
It is possible that there is a significant depth effect for nebulae towards Gem OB1.

Fig.~\ref{spiral} shows the angular distribution of the confirmed clusters in the Galactic plane and spiral arms \citep{Momany06}. A significant concentration of new clusters occurs along the Perseus arm.

\section{Concluding remarks}
\label{sec:8}

The present work investigates the nature of 48 overdensities from the catalogue of FSR07, projected towards the Galactic anticentre. Besides the 6 previously studied OCs, we confirm 18 of them as new clusters. Of the remainder, 6 are previously studied OCs, 7 are probable clusters, and 17 overdensities still require deeper photometry to check if they are clusters or plain field fluctuations.

In addition, we analyse the previously studied clusters Stock 8, Kronberger 1, and BPI 14. We discovered 7 clusters (CBB 3 to CBB 9). These objects together with FSR 780 and FSR 777 are located in the Aur OB2 association. This association presents a family of young clusters with ages younger than 10 Myr. On a smaller scale, the aforementioned objects (except CBB 8 that is embedded in Sh2-229) form an association of clusters with evidence of sequential star formation similar to Sh2-235 \citep{Camargo11}. Based on the distance derived for them, we argue that Aur OB2 is located in the Perseus arm at a distance of 2.7 kpc from the Sun. 

Aur OB2 may be a fundamental laboratory to investigate star formation related to sub-structured clusters and associations of clusters as well as the origin of field OB stars.

The confirmed clusters FSR 888 and FSR 890 are embedded in Sh2-249 inside Gem OB1. The distance derived for these ECs suggest that Sh2-249 and Gem OB1 are also objects of the Perseus arm with a distance of about 2.6 kpc. 

In total, we analysed 58 objects, deriving fundamental parameters for 28 and structural parameters for 13 of them. Most of the confirmed clusters are very young and located in the Perseus arm. In this sense, the present results represent a significant increase in the number of young clusters, especially in the Perseus arm.

The present work shows that, to uncover the nature of stellar overdensities, it is crucial to decontaminate for field stars. In particular, this procedure suggests that embedded clusters and/or PMS clusters are very common. We propose a conceptual separation of young clusters into actual ECs that are still embedded in nebulae, and those with PMS stars but essentially gas/dust free, as a consequence of evolutionary effects.

\vspace{0.8cm}
\textit{Acknowledgements}:
We thank an anonymous referee for constructive comments and suggestions.
This work was partially supported by CNPq and CAPES (Brazil).

\label{lastpage}
\end{document}